\documentclass[%
 reprint,
 amsmath,amssymb,superscriptaddress,
 aps,
showpacs]{revtex4-1}
\usepackage{graphicx}
\usepackage{caption}
\usepackage{color}
\usepackage{amsmath}
\usepackage{amssymb}
\usepackage{comment}
\usepackage{hyperref}
\usepackage{float}
\usepackage{setspace}
\usepackage{subfig}
\usepackage{tikz}
\usetikzlibrary{shapes,arrows}
\usepackage{titlesec}
\def\k{{{\bf k}}}

\def\Im{{\mbox{Im}}}

\definecolor{scarred}{rgb}{0.75,0.0,0.0}

\makeatletter
\renewcommand\@makecaption[2]{%
  \par
  \vskip\abovecaptionskip
  \begingroup
   \small\rmfamily
    \begingroup
     \samepage
     \flushing
     \let\footnote\@footnotemark@gobble
     \@make@capt@title{#1}{#2}\par
    \endgroup
  \endgroup
  \vskip\belowcaptionskip
}
\makeatother

\begin{document}
\title{Magnetization, d-wave superconductivity and non-Fermi liquid behavior in a crossover from dispersive to flat bands}
\author{Pramod Kumar}
\affiliation{Department of Applied Physics, Aalto University,
Helsinki, Finland}
\author{Tuomas I. Vanhala}
\affiliation{ Department of Physics,
Arnold Sommerfeld Center for Theoretical Physics (ASC),
Munich Center for Quantum Science and Technology (MCQST),
Fakult\"{a}t f\"{u}r Physik, Ludwig-Maximilians-Universit\"{a}t M\"{u}nchen,
M\"{u}nchen, Germany}
\author{P\"aivi T\"orm\"a}
\affiliation{Department of Applied Physics, Aalto University,
Helsinki, Finland}
\begin{abstract}
 We explore the effect of inhomogeneity on electronic properties of the two dimensional Hubbard model on a square lattice using dynamical mean field theory (DMFT). The inhomogeneity is introduced via modulated lattice hopping such that in the extreme inhomogeneous limit the resulting geometry is a Lieb lattice, which exhibits a flat-band dispersion. The crossover can be observed in the uniform sublattice magnetization which is zero in the homogeneous case and increases with the inhomogeneity. Studying the spatially resolved frequency dependent local self-energy, we find a crossover from Fermi-liquid to non-Fermi-liquid behavior happening at a moderate value of the inhomogeneity. This emergence of a non-Fermi-liquid is concomitant of a quasi-flat band. For finite doping the system with small inhomogeneity displays d-wave superconductivity coexisting with incommensurate spin-density order, inferred from the presence of oscillatory DMFT solutions. The d-wave superconductivity gets suppressed for moderate to large inhomogeneity for any finite doping while the incommensurate spin-density order still exist.
\end{abstract}
\pacs{ Strongly correlated electron systems, Non-Fermi-liquid ground state, Cold atoms}
\maketitle
\section{Introduction}
\label{sec:intro}

In his famous 1989 paper \cite{PhysRevLett.62.1201}, Lieb considered Hubbard models on certain bipartite lattices with highly degenerate single-particle states, which he showed to have ground states with nonzero spin. Magnetism in such flat-band models has later been the subject of many theoretical and computational studies
\cite{PhysRevLett.69.1608,Mielke1993,doi:10.1143/PTP.99.489,PhysRevLett.88.127202,
  PhysRevA.80.063622,PhysRevA.90.043624,PhysRevB.93.235143,PhysRevB.94.155107}. Another
type of order studied in connection with flat bands is
superconductivity, where electronic pairing can be enhanced by the
high density of states
\cite{shaginyan1990superfluidity,PhysRevB.83.220503,PhysRevB.90.094506,Peotta2015}. While
flat-band models such as the Lieb lattice were originally intended as
theoretical toy models, developments in experimental techniques in
ultracold gases and condensed matter systems now allow them to be
created and studied. A striking example of how flat bands can enhance
correlation effects is twisted bilayer graphene
\cite{Cao2018,Cao20181,Yankowitz1059}, where certain ``magic'' twist
angles lead to superconductivity and insulating states whose precise nature
is not yet understood.
Flat-band systems have also been engineered by
manipulating the electronic surface states of a copper crystal using
adsorbed molecules \cite{Slot2017,Drost2017,Leykam2018} and using
optical potentials for bosonic
\cite{Taiee1500854,PhysRevLett.118.175301,Leykam2018} and fermionic
\cite{taie2017spatial} ultracold quantum gases. The advantage of these
experiments is the high degree of tunability in the lattice
parameters.

In both electronic and optical lattice experiments the most commonly
used flat-band model system is the Lieb lattice \cite{Leykam2018},
whose simple structure makes it relatively easy to be
implemented. However, because of experimental imperfections, exactly
flat bands are difficult to achieve. This motivates us to introduce a
model, pictured in Fig. \ref{fig1:band structure}, that is an
interpolation between the Lieb lattice and the simple square lattice,
and exhibits a band with a \emph{tunable} bandwidth. This can also be
compared to twisted bilayer graphene where the width of the low-energy
bands can be tuned by changing the twist angle
\cite{Bistritzer12233}. A related idea where suitably chosen
next-nearest-neighbour hoppings lead to partially flat bands and
typical flat-band effects such as enhanced superconducting transition
temperatures and non-fermi-liquid behaviour, has been studied in
recent works \cite{PhysRevB.99.235128,2019arXiv190309888S}. A
$\pi$-flux lattice model \cite{PhysRevB.97.235152} exhibiting Dirac
fermions with a tunable velocity has also been considered. Our main
goal here is to study the crossover between the flat-band physics and
normal dispersive behaviour, allowing us to build a general picture of
how flat-band effects on magnetic states would be observed in
experiments. Interestingly, the model also provides a new perspective to flat-band
ferromagnetism on the Lieb lattice: We find that the ground state as a
whole is always antiferromagnetic with no total magnetic
moment. However, as the model is tuned towards the Lieb lattice limit,
a subset of the lattice sites carrying a magnetic moment becomes
weakly coupled to the rest of the lattice. Thermal fluctuations easily
reduce the magnetization of the weakly coupled part, thus leading to a
total magnetization that \emph{increases} with temperature.

Another motivation for our work is to study how the $d$-wave
superconducting states of the square lattice model
\cite{vanhala2017dynamical} interact with the flat band. While the
general idea is that flat bands can boost interaction effects by
decreasing the competition from kinetic energy, leading to strong
correlations and high critical temperatures for ordered states, this
is not the whole story: For ordered states resulting from strong correlation effects beyond
the mean-field level, the single-particle band structure may be rather
irrelevant. In fact, we show that the $d$-wave superconductivity
is monotonically suppressed as the model is tuned towards the Lieb-lattice limit,
which is apparently because the asymmetry between the $A$ and $D$ sites
is incompatible with the local, correlated mechanism leading to the $d$-wave
pairing. In this context the model is best seen as a type of
\emph{inhomogeneous} square-lattice model, meaning a
model where the hopping amplitudes or on-site potentials can vary spatially.
Motivation for such models
is related to the so-called stripe order, i.e. spatially non-uniform
spin-density or charge-density order, which has been found in several
families of the cuprates~\cite{Keimer2015,Zheng1155} and also in
ultra-cold atom experiments recently~\cite{salomon2018}, albeit only
in 1D systems. Other examples of inhomogeneity include quasi-periodic
systems~\cite{PhysRevLett.53.1951}, fermionic ultra-cold atoms in
harmonic traps~\cite{RevModPhys.80.885}, electron systems on
surfaces~\cite{PhysRevB.59.R2474}, interfaces~\cite{Ohtomo2002} and
topological insulating
systems~\cite{Hohenadler_2013,PhysRevB.94.115161}.

Whether the presence of incommensurate spin and density order competes
with or helps the emergence of superconductivity (SC)  in real materials is in general
unsettled~\cite{Fradkin2012,Julien914}. Theoretical studies report both
suppression and enhancement of dSC order with inhomogeneity
~\cite{PhysRevB.81.214525,PhysRevB.90.075121,PhysRevB.78.020504,PhysRevB.84.054545,PhysRevB.77.214502,PhysRevB.90.075121,Wu_2019}, depending on the inhomogeneity pattern and strength, interaction strength and doping of the system.
%In a few theoretical studies an enhancement of the dSC order with %inhomogeneity~\cite{PhysRevB.81.214525,PhysRevB.90.075121} has been
%found, while others report
%suppression~\cite{PhysRevB.78.020504,PhysRevB.84.054545}.
%\textcolor{blue}{
%In fact, finding the best strength and pattern of inhomogeneity for
%superconductivity can be seen as an optimization problem, and
%the result depends e.g. on the interaction strength~\cite{PhysRevB.77.214502,PhysRevB.90.075121}.}
A much studied inhomogeneity pattern is the 2D Hubbard model on a
checkerboard lattice where the strong and weak nearest neighbor
hopping amplitudes alternate along both
directions~\cite{Wu_2019,PhysRevB.77.214502,PhysRevB.78.020504,PhysRevB.88.214518,PhysRevB.76.161104}. A stripe version of the model,
where the nearest neighbor hopping amplitude is modulated along one
direction, has also been considered~\cite{PhysRevB.77.214502}. Other
inhomogeneity patterns are checkerboard- and stripe-like variations in
the local onsite potential on $2 \times 2$
plaquettes~\cite{PhysRevB.81.214525}. 
To study the inhomogeneous square-lattice Hubbard model introduced
here, we employ dynamical mean-field theory (DMFT) and its cluster
extensions. In section~\ref{sec:model}, we introduce the model as an
interpolation between the square and the Lieb lattice, followed by the
formalism of real space DMFT to capture spatially resolved local order
parameters and cellular DMFT that can capture the non-local
correlations essential to dSC. In sections~\ref{sec:magnetism}
and~\ref{sec:double}, we discuss the effect of the inhomogeneity and
the quasi-flat band on the emergent magnetic order and the double
occupancy, respectively. The breakdown of the Fermi-liquid behavior in
the crossover from dispersive to flat-band behaviour is discussed in
section~\ref{sec:nfl}.  Finally, we discuss the effect of the
inhomogeneity on the behavior of dSC and incommensurate spin- and
density-wave order in section~\ref{sec:dop}.

\section{Model and method}

\label{sec:model}
\begin{figure}
\includegraphics[scale=0.45]{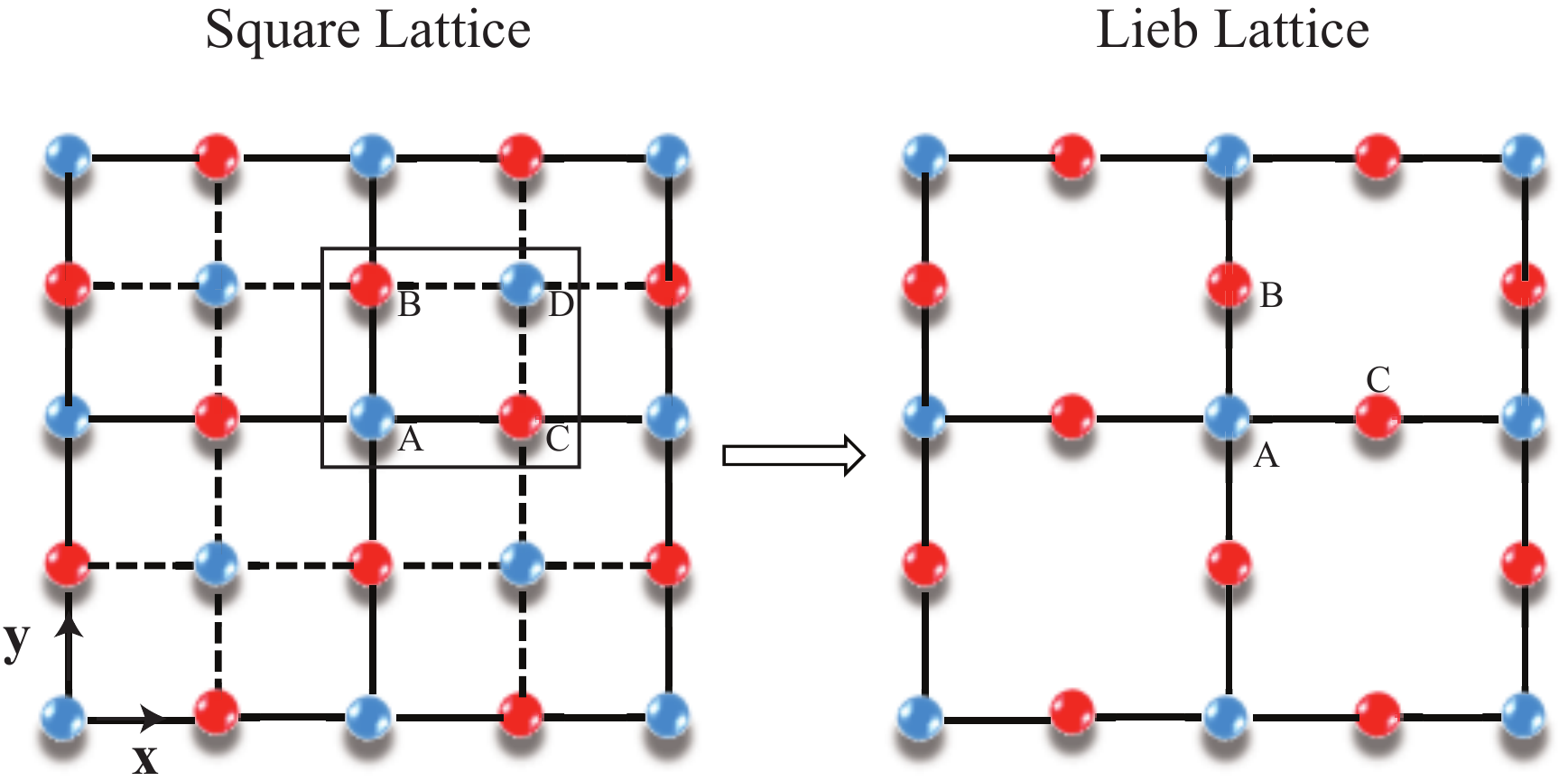}\\
\includegraphics[scale=0.5]{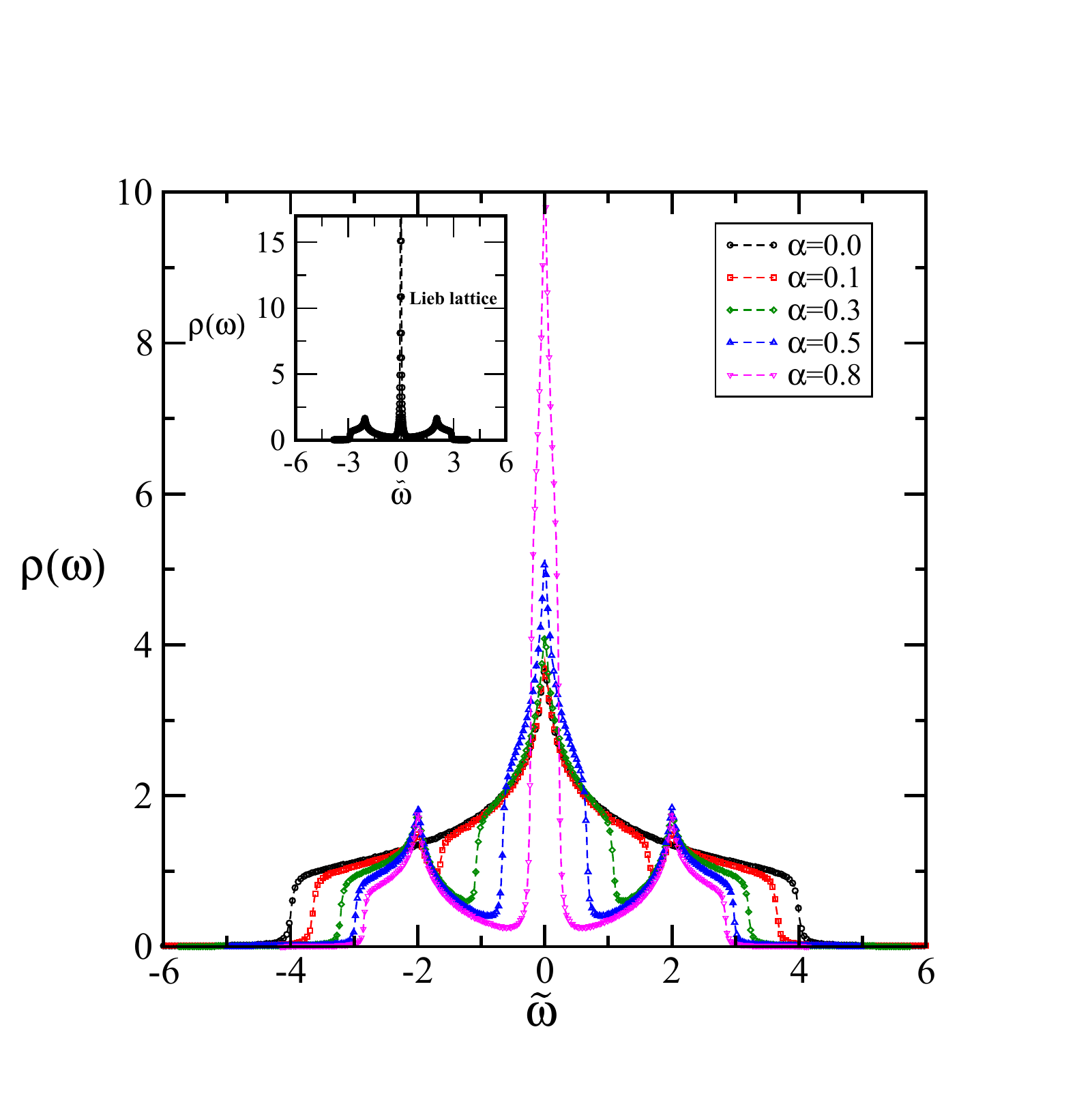}

\caption{Upper panel: Schematic  representation of the inhomogeneity introduced by modulated hopping. The solid lines represent the hopping amplitude $(1+\alpha) t$, while the dashed lines represent $(1-\alpha) t$. The square drawn by the solid line is the smallest possible unit cell that captures magnetic and superconducting order parameters emerging at finite Hubbard interaction. In the limit $\alpha = 1$, the square lattice with modulated hopping turns into the Lieb lattice.
Lower panel: The non-interacting density of states (DOS) of the inhomogeneous lattice as a function of the normalized energy parameter $\tilde{\omega}$ (see text) for different choices of $\alpha$. The density of states evolves from square lattice behavior to Lieb lattice one. In the inset: DOS for the pure Lieb lattice.}
\label{fig1:band structure}
\end{figure}
The grand canonical Hamiltonian of the Hubbard model for an inhomogeneous square lattice as shown in Fig.~\ref{fig1:band structure}(a) can be expressed as $H=H_{\text{t}}-\mu N+H_U$, where the first term is the tight-binding part represented in standard second quantized notation as
\begin{align}
H_{\text{t}}=&-\sum_{\langle ij \rangle,\sigma}\Big[(t_{ij} c_{i,\sigma}^\dagger c^{\phantom{\dagger}}_{j,\sigma}+ \text{h.c.}) \Big] \label{eq1},
\end{align}
where $c_{j,\sigma}^\dagger$ is the creation operator corresponding to different sites of the unit cell at $j = (x, y)$ and $\sigma$ labels spin. We have introduced the inhomogeneity via the modulated next-nearest hopping by setting $t_x=t(1+(-1)^y\alpha)$ and $t_y=t(1+(-1)^x\alpha)$, where $\alpha=0$ corresponds to the homogeneous square lattice and $\alpha=1$ represents the Lieb lattice as shown in Fig.~\ref{fig1:band structure}. The Lieb lattice resembles the CuO$_2$ planes of the high-$T_c$ cuprate superconductors with blue(red) circles representing Copper(Oxygen) ions~\cite{PhysRevB.90.094506}. However, there is huge on-site energy difference between the Copper and Oxygen orbitals, violating one of the criterion of Lieb's theorem~\cite{PhysRevLett.62.1201}. Additionally, the ground state of the cuprates is aniferromagnetic unlike the ferromagnetic ground state of the Lieb lattice. The second term $\mu N$ of the full Hamiltonian introduces the chemical potential, where the total particle number is $N=\sum\limits_{ j,\sigma } c_{j,\sigma}^\dagger c^{\phantom{\dagger}}_{j,\sigma}$. 

The last term is the on-site Hubbard interaction which can be defined as
\begin{equation}
H_U=U\sum\limits_{j} (n_{j,\uparrow}-\frac{1}{2})(n_{j,\downarrow}-\frac{1}{2}),
\label{eq_Hu}
\end{equation}
where $U$ is the interaction strength with $U>0$ for the repulsive Hubbard model. To account for the  aforementioned inhomogeneity, the smallest possible unit cell has four sites as shown by the solid square in Fig.~\ref{fig1:band structure}. The tight-binding Hamiltonian in momentum space can be written as
 \begin{align}
H_{\text{t}}=&\sum_{k,\sigma} \psi_{\k \sigma}^\dagger H_{\text{t}}(\k) \psi_{\k \sigma} \nonumber,
\end{align}
where $\psi_{\k\sigma}=(c_{A\sigma} \  c_{B\sigma} \ c_{C\sigma}\ c_{D\sigma})^T$ and
$$
H_{t}(\k)=
-2\begin{pmatrix} 
0 & t_{+}\cos k_x & t_{+}\cos k_y & 0 \\
t_{+}\cos k_x&  0 &0 & t_{-}\cos k_y \\
t_{+}\cos k_y & 0 & 0 & t_{-}\cos k_x \\
0 & t_{-}\cos k_y& t_{-}\cos k_x & 0
\end{pmatrix}
$$
with $t_{\pm}=(1\pm\alpha)t$ . The energy eigenvalues of the tight binding Hamiltonian can be given as
\begin{equation}
E_{\k}=\pm 2t\sqrt{(1+\alpha^2)S_{+}\pm\sqrt{(1+\alpha^2)^2S_{+}^2-(1-\alpha^2)^2S_{-}^2}},
\end{equation}
where $S_{+}=\cos^{2}k_x+\cos^{2}k_y$ and $S_{-}=\cos^{2}k_x-\cos^{2}k_y$. For  $\alpha=1$, the resulting geometry is the Lieb lattice with $E_{\k}=0$ and $E_{\k}=\pm2\sqrt{2}t\sqrt{\cos^2 k_x+\cos^2 k_y}$ .
In the lower panel of the figure~\ref{fig1:band structure}, we show the DOS vs $\tilde{\omega}$, where $\tilde{\omega}=\omega/t(1+\alpha)$, for the tight-binding part of the Hamiltonian for different choices of $\alpha$. The DOS has a Van-Hove singularity at zero energy for $\alpha=0$, which grows with increasing inhomogeneity parameter $\alpha$ into a narrow peak structure, ultimately turning into a $\delta$- function representing the flat band of the Lieb lattice for $\alpha=1$ ( see inset of Fig.~\ref{fig1:band structure}).

To investigate the effects of correlations and inhomogeneity at half-filling, we have employed real-space dynamical mean-field theory (RDMFT) \cite{1367-2630-10-9-093008,PhysRevLett.113.185301} for finite Hubbard interactions. DMFT maps a lattice problem into an effective single impurity problem taking into account the lattice effects in a self-consistent manner~\cite{RevModPhys.68.13}.
Within single-site DMFT the self-energy $\Sigma_{ij\sigma}(i \omega_n)$ is assumed to be spatially local and uniform, so that $\Sigma_{ij\sigma}(i \omega_n) \sim \delta_{ij}\Sigma_{\sigma}(i \omega_n)$. The $i$ and $j$ index the lattice sites, $\omega_n=\pi(2n+1)T$, where $T$ is the temperature, are the Matsubara frequencies and $\sigma$ is the spin index. For the inhomogeneous  case, however, the uniformity assumption is relaxed. Hence we use RDMFT where the self-energy is still local but varies spatially, i.e. $\Sigma_{ij\sigma}(i\omega_n)=\Sigma^{i}_{\sigma}(i\omega_n)\delta_{ij}$~\cite{1367-2630-10-9-093008}. 
  
The RDMFT method for a given unit cell can be described as follows. The local Green's function of the lattice system can be calculated as
\begin{equation}
  \mathbf{G}_\sigma(i\omega_n)= \frac{1}{N_{\mathbf{k}}}\sum_{\mathbf{k}} \left( \mathbf{G}_{\mathbf{k} \sigma}^0(i\omega_n)^{-1}-\mathbf{\Sigma}_\sigma(i\omega_n) \right)^{-1}, \label{eq:dmft}
\end{equation}
where the bold quantities are matrices of the dimension $4
\times4$ and $N_\mathbf{k}$ is the number of $\mathbf{k}$- points. Thus the matrix element $\mathbf{G}_{\sigma}(i \omega_n)_{ij}$ is the Green's function between sites $i$ and $j$ of the unit cell. The non-interacting Green's function $\mathbf{G}_{\mathbf{k} \sigma}^0(i\omega_n)^{-1}=\mu_\sigma+i\omega_n-\mathbf{T}_{\mathbf{k}}$, where $\mathbf{T}_{\mathbf{k}}$ is the superlattice Fourier transform of the hopping matrix. The self-energy is assumed to be diagonal in the site indices. For each site $i$ in the unit cell, there is an effective single impurity Anderson model, which is defined by the dynamical Weiss mean-field
\begin{equation}
\mathcal{G}_{\sigma}^{i}(i\omega_n)^{-1}=(\mathbf{G}_{\sigma}(i\omega_n)_{ii})^{-1}+\Sigma^{i}_{\sigma}(i\omega_n)_{ii}.\label{eq:weis}
\end{equation}
Using the Weiss function $\mathcal{G}_{\sigma}^{i}$, we calculate the self-energy of each of the impurity problems using an impurity solver. These new self-energies are supplied again to equation \ref{eq:dmft} and the process is iterated to find a converged solution.

We use exact diagonalization (ED) and continuous time quantum Monte Carlo (CT-INT) as impurity solvers at zero temperature and finite temperature, respectively.~\cite{vanhala2017dynamical,PhysRevB.76.035116}. We define the local magnetization, $m_{i}=n_{i,\uparrow}-n_{i,\downarrow}$, where $n_{i,\sigma}=G_{i,\sigma}(\tau \rightarrow 0^-)$ is the density of spin-$\sigma$ particles for a given site of the unit cell. Another important quantity to measure the effects of correlation is the double occupancy $D=\langle n_{i,\uparrow}n_{i,\downarrow} \rangle$, representing the tendency of two particles to occupy the same site. It is $0.25$ in the zero interaction limit while it vanishes in the Mott insulating large $U$ limit for the repulsive Hubbard model for a homogeneous system at half-filling. It can be directly calculated using DMFT+CT-INT as
\begin{equation}
D=\frac{n_i}{2}-\frac{\langle k \rangle_{\text{MC}}}{\beta |U|}-\frac{1}{4}
\end{equation}
where $n_i=n_{i,\uparrow}+n_{i,\downarrow}=1$ for half-filling and $k_{\text{MC}}$ is the Monte-Carlo perturbation order~\cite{PhysRevB.76.035116}. Additionally, the double occupancy for a site can be directly compared with the local moment $m_i^2$ measured in the experiments~\cite{Jordens2008}, given as
\begin{equation}
\langle m_i^2 \rangle =1-2\langle n_{i\uparrow}n_{i\downarrow}\rangle\label{eq13}.
\end{equation}

 To study superconductivity within DMFT, we use the Nambu formalism~\cite{RevModPhys.77.1027,PhysRevB.84.054545}, where the Green's function can be written in the Nambu-spinor notation as
\begin{eqnarray}
G_{ij}(\tau)=-\langle \mathcal{T} \psi_{i}(\tau) \psi_{j}^\dagger(0)\rangle,
\end{eqnarray}
where $\psi_i(\tau)\equiv (c_{i\uparrow}, c_{i\downarrow}^\dagger)^T$ and its matrix notation can be given as
$$
\mathbf{G}(\tau)=
\begin{pmatrix} 
\mathbf{G}_{\sigma}(\tau) & \mathbf{F}(\tau) \\
\mathbf{F}^{\dagger}(\tau) & -\mathbf{G}_{\bar{\sigma}}(-\tau)
\end{pmatrix},
$$
where $\tau$ is imaginary time, $G_{ij\sigma}(\tau)\equiv-\langle \mathcal{T} c_{i\sigma}(\tau) c_{j\sigma}^\dagger(0)\rangle$ and $F_{ij}(\tau)\equiv-\langle \mathcal{T} c_{i\downarrow}(\tau) c_{j\uparrow}(0)\rangle$ are the normal and anomalous Green's functions, respectively.
To capture a non-local dSC order parameter, emerging away from half-filling, we employ cellular dynamical mean field theory (CDMFT). Within CDMFT, a lattice problem is mapped to a finite cluster coupled to a non-interacting bath. In our case the cluster is a four site $(2\times2)$ plaquette as shown in Fig.~\ref{fig1:band structure}, which has been used to study the dSC order in the canonical square lattice Hubbard model \cite{vanhala2017dynamical,PhysRevB.62.R9283,PhysRevB.74.054513}. The local cluster Green's function of the lattice system is given by the matrix equation
\begin{equation}
\mathbf{G}_c(i\omega_n)= \frac{1}{N_{\mathbf{k}}}\sum_{\mathbf{k}} \left( \mathbf{G}^0(\mathbf{k},i\omega_n)^{-1}-\mathbf{\Sigma}_c(i\omega_n) \right)^{-1}, \label{eq4}
\end{equation}
where $N_{\mathbf{k}}$ is the number of $\mathbf{k}$- points. The non-interacting Green's function $\mathbf{G}^0(\mathbf{k},i\omega_n)^{-1}=i\omega_n+\mu\sigma_z-\mathbf{T}(\mathbf{k})\sigma_z$, where $\mathbf{T}_{\mathbf{k}}$ is the super-lattice Fourier transform of the hopping matrix with dimension equal to the number of sites in the cluster i.e. $4\times4$. The cluster self-energy $\mathbf{\Sigma}_c(i\omega_n)$ can be given as 
$$
\mathbf{\Sigma}_c(i\omega_n)=
\begin{pmatrix} 
\mathbf{\Sigma}_{\uparrow}(i\omega_n) & \mathbf{S}(i\omega_n) \\
\mathbf{S}(i\omega_n) & -\mathbf{\Sigma}_{\downarrow}^{*}(i\omega_n)
\end{pmatrix}
$$
where $\Sigma_{ij\sigma}( i\omega_n)$ $(S_{ij}(i\omega_n))$ is the normal (anomalous) part of the self-energy matrix of  dimension $4\times4$.

 Similar to the the single site DMFT, there is an effective impurity problem for the cluster, which can be defined by the Weiss mean-field
\begin{equation}
\bf{\mathcal {G}}^0_c(\textit{i}\omega_{\textit{n}})^{-1}=\mathbf{G}_c^{-1}(\textit{i}\omega_{\textit{n}})+\mathbf{\Sigma}_{c}(\textit{i}\omega_{\textit{n}}).
\end{equation}
This quantity is also known as the "bath function", and represents the non-interacting Green's function of the impurity problem. Given the mean-field $\bf{\mathcal {G}}^0_c$, we calculate the cluster propagator and the self-energy, $\Sigma_c(i\omega_n)$ from the above Weiss mean-field using ED as an impurity solver~\cite{vanhala2017dynamical}. The process is iterated similar to single site DMFT to find the solution. We define the average magnetization for the cluster as
\begin{equation}
m_{\text{avg}}=\sum_i\frac{|m_i|}{4}, \label{eq6}
\end{equation}
where $m_i$ is the local magnetization of a given site calculated from the normal Green's function. Additionally, we can define the average dSC order for the given four site cluster as
\begin{equation}
\Delta_{\text{avg}}=\frac{\Delta_{\text{AB}}+\Delta_{
\text{BD}}+\Delta_{\text{DC}}+\Delta_{\text{CA}}}{4},\label{eq:dwave}
\end{equation} 
$\Delta_{ij}= S_{ij}(\langle c_{i\downarrow} c_{j\uparrow}\rangle-\langle c_{i\uparrow} c_{j\downarrow}\rangle )/ 2$ with $\langle c_{i\downarrow} c_{j\uparrow}\rangle=F_{ij}(\tau \rightarrow 0^{-})$ and $F_{ij}(\tau)$ is the non-local anomalous Green's function of the unit cell. For the singlet $d_{x^2-y^2}$ pairing  on a square lattice we have
\[
    S_{ij}=\left\{
                \begin{array}{ll}
                  \ \ 1 \ \ \text{if} \ \ i-j=\pm\hat{x}\\
                  -1 \ \ \text{if} \ \ i-j=\pm\hat{y}
                \end{array}
              \right.
  \]
  where $\hat{x}$ and $\hat{y}$ are unit lattice vectors.

\section{Results}
In this section, we discuss the effect of the inhomogeneity and Hubbard interaction for two cases: 1) half-filling, where number of particles per site is one, and 2) away from half-filling with finite doping $x=1-n_{\text{avg}}$, where $n_{\text{avg}}=\sum_{i}\frac{n_i}{4}$ is the average density over the unit cell. At half-filling, the interplay of inhomogeneity and interaction is visible in the local magnetism and the double occupancy. One of the key purposes of this work is to study the quasi-particle behavior in the inhomogeneous system. We calculate the local self-energy and show breakdown of Fermi-liquid behavior with increasing strength of the inhomogeneity. The result that we find can be associated to the (quasi-)flat band present in the inhomogeneous system. Away from half-filling, we present a phase diagram as a function of inhomogeneity, $\alpha$, and chemical potential, $\mu$,  showing $\Delta(\mu, \alpha)$ at $U=6.0$. We also show the averaged magnetic order, $m_{\text{avg}}$, for different set of $\mu$ and $\alpha$. The dSC order parameter decreases with $\alpha$ and vanishes further for moderate to large $\alpha$. The magnetic order coexists with dSC for small values of $\alpha$. For a moderately inhomogeneous system, incommensurate magnetic order is present with no dSC.
\subsection{Magnetism}
\label{sec:magnetism}
\begin{figure}
\includegraphics[scale=0.6]{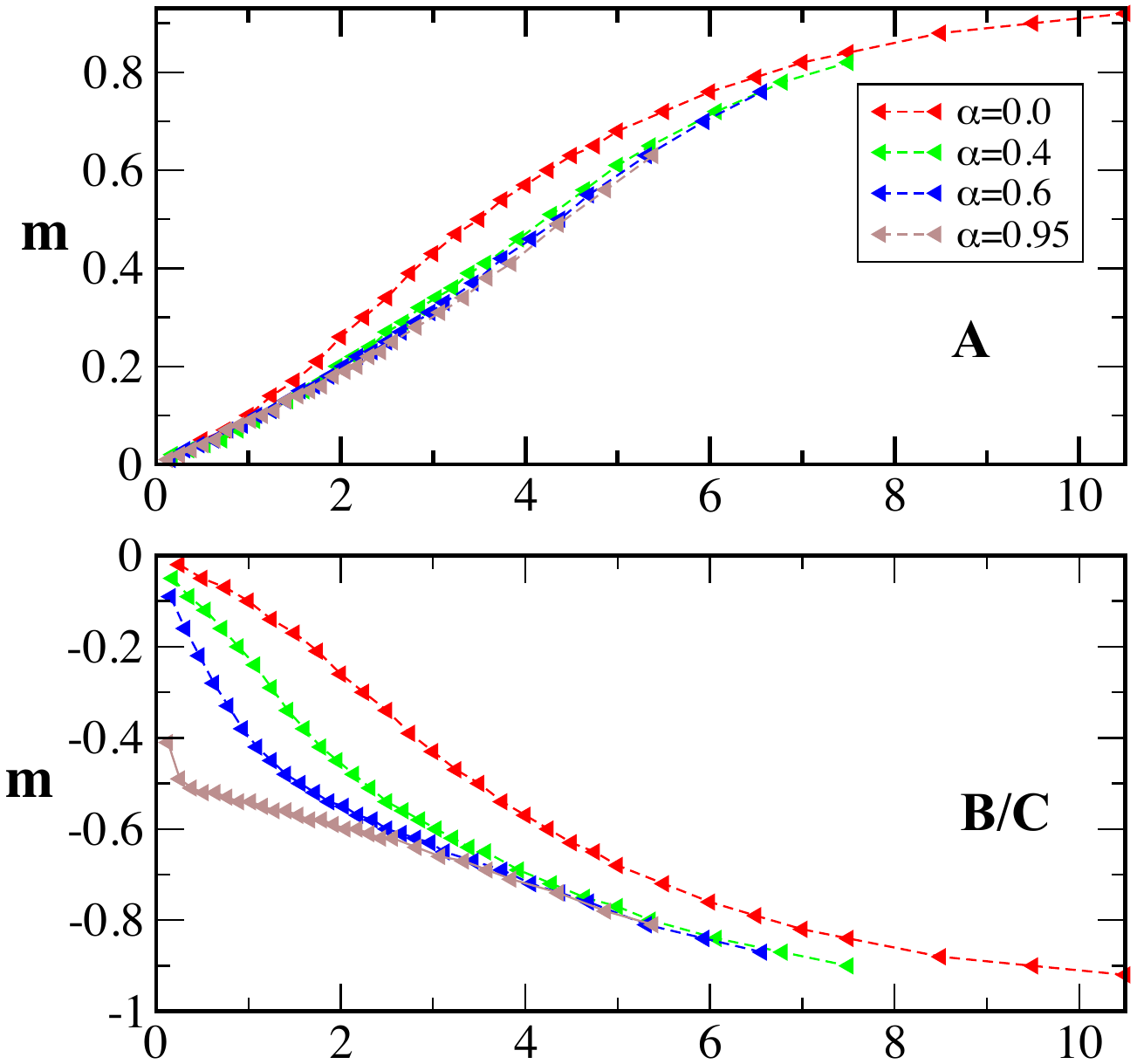}
\includegraphics[scale=0.6]{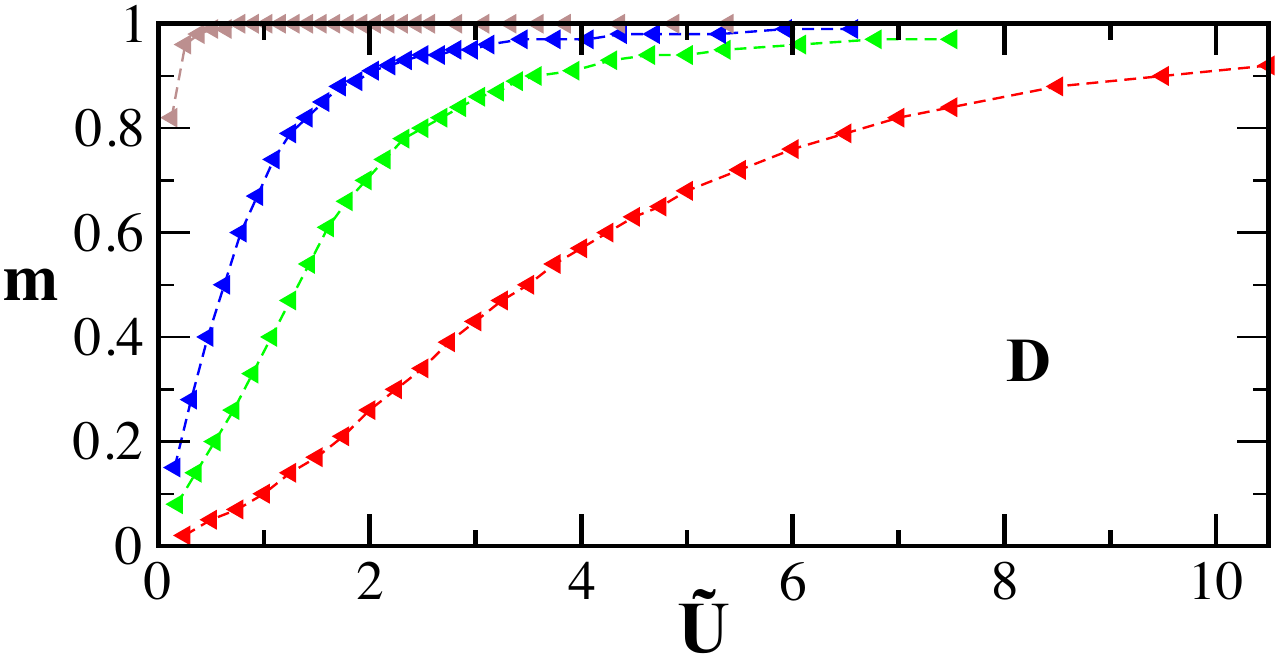}

\caption{Magnetic order parameter $m$ for $A$ site (upper panel), $B$ and $C$ sites (middle panel), and $D$ site (lower panel) for varying $\tilde{U}$ and different inhomogeneity $\alpha$ at zero temperature. The Hubbard interaction $U$ has been scaled by  $t_+=(1+\alpha)t$.}
\label{fig2:magnetism}
\end{figure}

Due to the spatial inhomogeneity introduced by the modulated hoppings, the local magnetic order is non-uniform across different sites. We show the spatially resolved magnetic order $m$ evaluated using ED+RDMFT at zero temperature for varying interaction strength $\tilde{U}=U/t_+$ at different $\alpha$ in figure~\ref{fig2:magnetism}. We allow the breaking of the $SU(2)$ spin-rotation symmetry to capture the magnetically ordered state. An initial self-energy that is constant in the Matsubara frequency is added in way that it breaks SU$(2)$ symmetery of the Hamiltonian. For a homogeneous system, i.e. $\alpha=0$, local magnetic order gradually develops with finite Hubbard interactions such that $m_A=-m_{B/C}=m_D$ for any $\tilde{U} > 0$. For weak interaction, the behavior of the magnetic order is consistent with Hartree-Fock mean-field theory~\cite{PhysRevB.31.4403} and saturates to unity in the Heisenberg limit for strong interactions. For any small but non-zero $U$, the absolute value of $m_{B/C}$ increases with increasing $\alpha$ such that $|m_{B/C}|~\sim 0.5$ for $\alpha \rightarrow 1$. Such finite local magnetization at $B/C$ sub-lattice for infinitesimal interaction  is caused by a flat band state with constant energy dispersion $E_k\approx 0$ located at the Fermi level. The high spin degeneracy is lifted already by the infinitesimal $U$, and magnetization develops at the $B/C$ sites that carry the flat band \cite{PhysRevB.96.245127}. For $\alpha\rightarrow 1$ the local magnetization at sub-lattice $D$ saturates to unity for infinitesimal $U$ since the $D$ sites get weakly coupled to the rest of the lattice. It is important to note that at $T=0$ the total magnetization summed over the unit cell is zero for $\alpha \in [0 \ 1)$ and finite $U$ although the absolute value of magnetization at different sites is different. For $\alpha=1$, sub-lattice $D$ gets isolated from the rest of the lattice showing zero local magnetization for any finite $U$ and we get a ferromagnetic ground state which is consistent with the Lieb theorem~\cite{PhysRevLett.62.1201}. At finite temperature thermal fluctuation suppresses the local magnetic order of the weakly coupled sub-lattice $D$ giving rise to a nonzero total magnetization summed over the unit cell also for $\alpha\rightarrow  1$. For the weakly interacting regime, the behavior of $m_{B/C}$ \textit{vs} $\tilde{U}$ changes from exponential to linear for $\alpha \sim 1$ due to the flat band~\cite{PhysRevB.96.245127}. Linear behavior of the order parameter with the Hubbard interaction in the weakly interacting regime can be explained with a simple mean-field gap equation with a $\delta-$ function density of states~\cite{PhysRevA.91.063610}. Local magnetizations for all sites coalesce to single curves for all values of $\alpha$ in the strong coupling regime, where the fermions are completely localized so that the system can be described by an effective Heisenberg model and the lattice geometry is insignificant to the behavior of local magnetic order.

\begin{figure}
\includegraphics[scale=0.75]{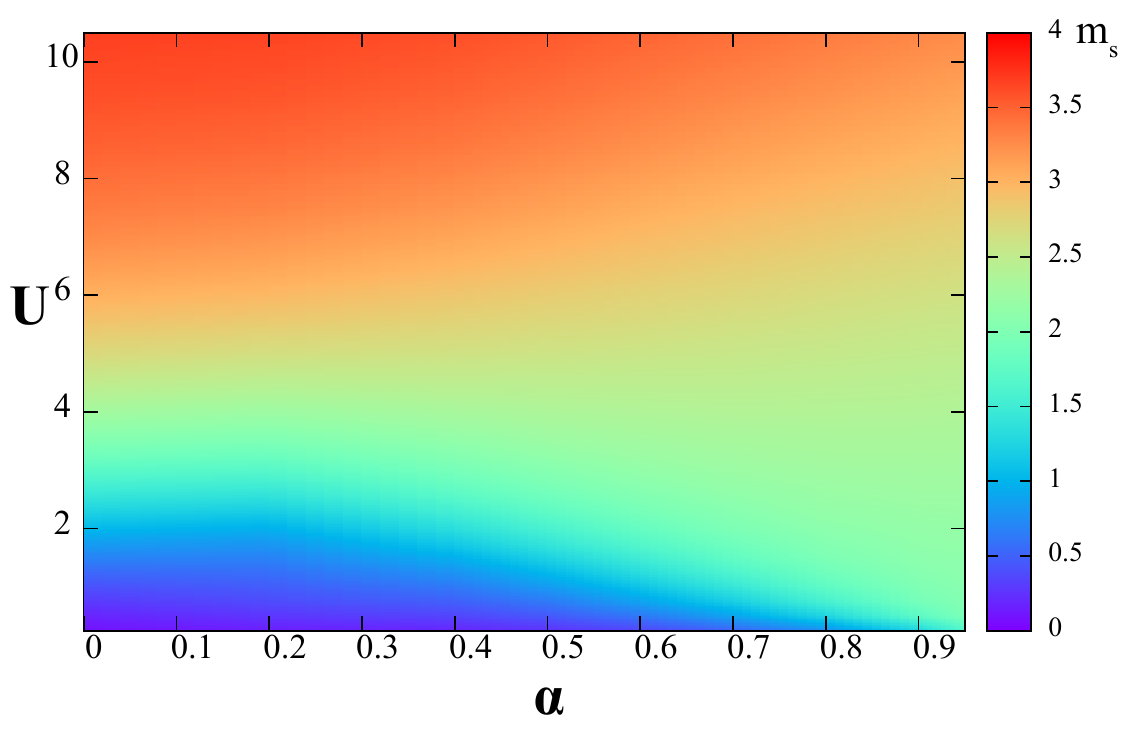}
\includegraphics[scale=0.75]{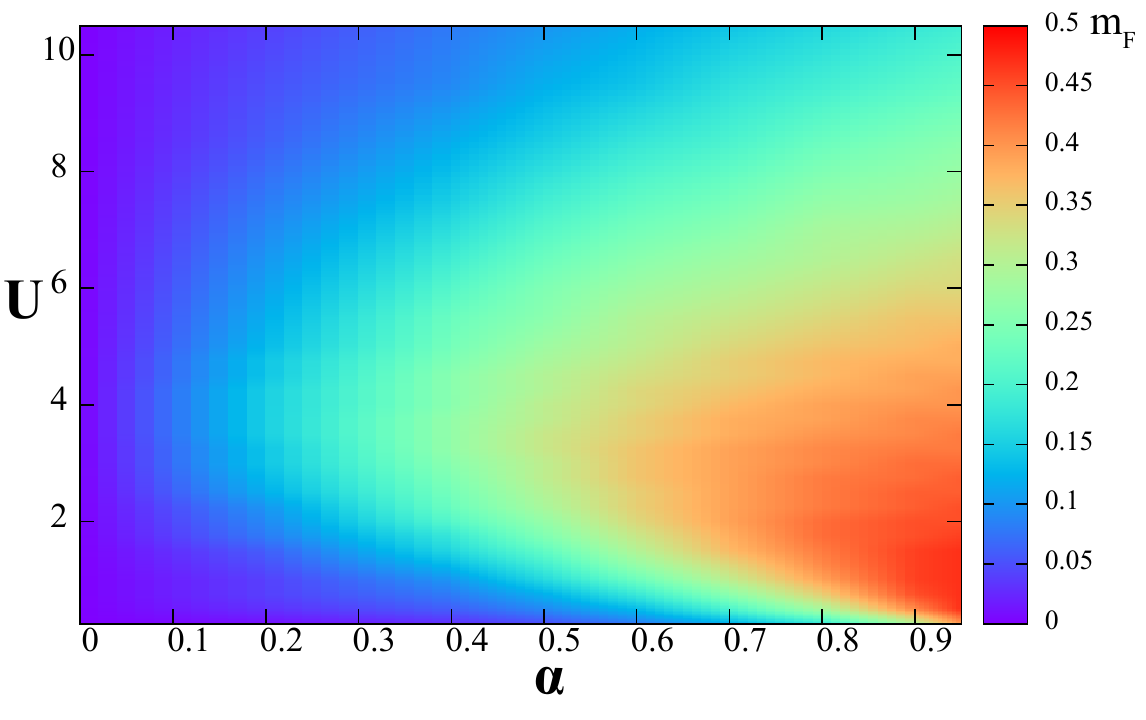}\caption{Upper panel: Phase diagram of the inhomogeneous Hubbard model showing staggered magnetization $m_s$ for different inhomogeneity parameter $\alpha$ and interaction $U$. Lower panel: Uniform magnetization for  the set of $\alpha$ and $U$. Here $m_{\text{F}}$ is maximal for $\alpha \rightarrow 1$ and $U\rightarrow 0$.}
\label{fig:3d staggered}
\end{figure}

\begin{figure}
\includegraphics[scale=0.6]{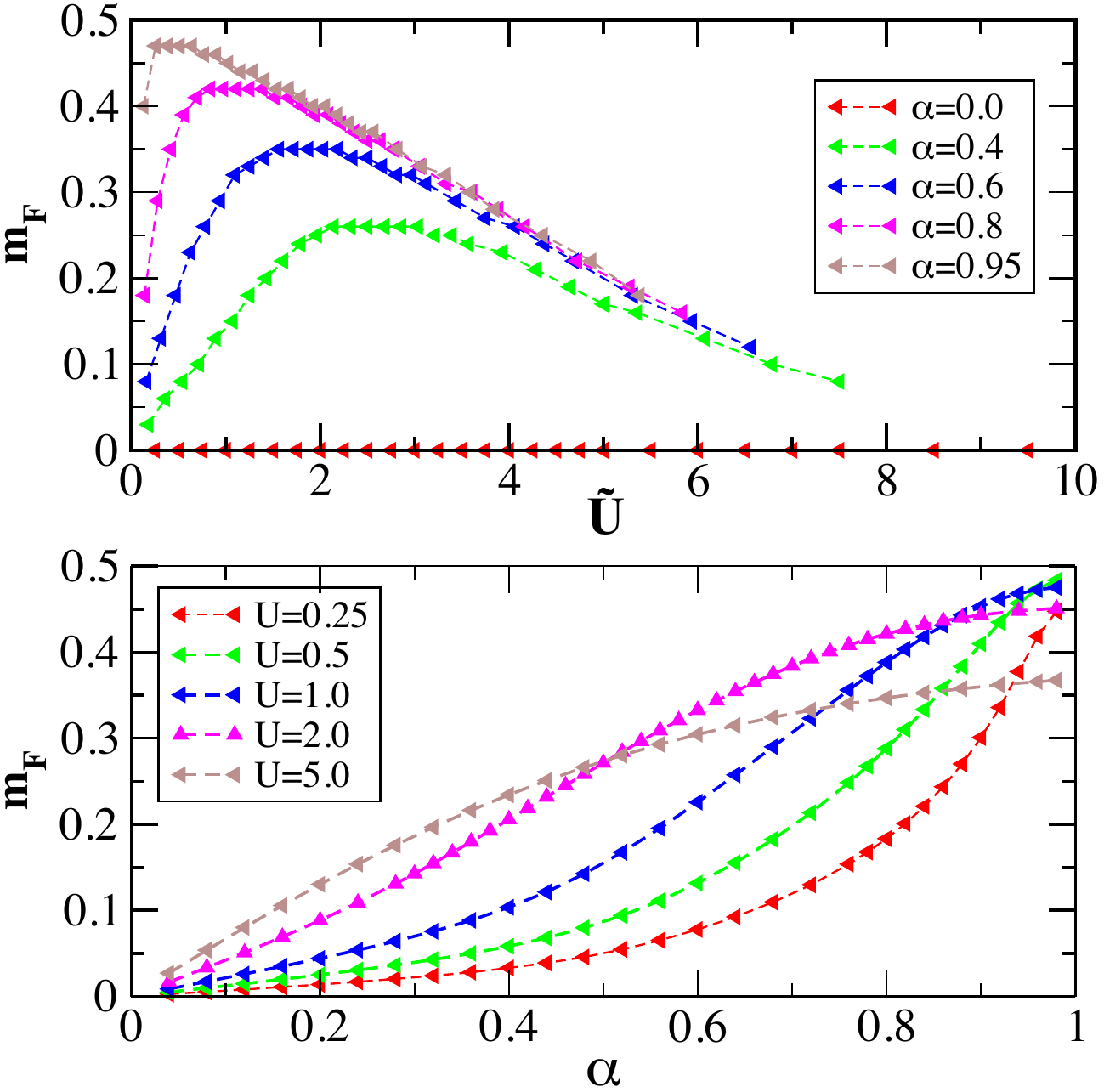}
\caption{Upper panel: Uniform magnetization $m_{\text{F}}$ for varying $\tilde{U}$ and different $\alpha$. Lower panel: $m_{\text{F}}$ for varying $\alpha$ for different  $U$.}
\label{fig3:staggered magnetism}
\end{figure}

 In the upper panel of figure~\ref{fig:3d staggered}, we show the phase diagram for staggered magnetization, i.e. $m_s=m_A + m_D-2 \ m_{B/C}$, obtained using ED+RDMFT for varying interaction $U$ and inhomogeneity $0 \le \alpha< 1$. For smaller interactions e.g. $U<2$, staggered magnetization assumes a finite value for moderate inhomogeneity such that $m_s\sim 2$ for $U\rightarrow 0^+$ and $\alpha\rightarrow 1$. In the strong coupling Heisenberg limit absolute value of the local magnetization at different sub-lattices asymptotically goes to unity for all inhomogeneities and thus $m_s\approx 4$, as evident from figure~\ref{fig2:magnetism}. In order to understand the effect of inhomogeneity on spatial distribution of the magnetic order, we show the behavior of uniform magnetization, i.e. $m_{\text{F}}=-(m_{\text{B/C}}+m_{\text{A}})$, in the lower panel of figure~\ref{fig:3d staggered}. The uniform magnetization, $m_F$, is zero for the homogeneous system for any finite interactions. Also $m_F$ is zero and independent of $\alpha$ in the strongly interacting regimes. However, it gets finite for moderate $\alpha$ and finite but moderate values of $U$. It has maximum value for $\alpha\rightarrow 1$ and $U\rightarrow 0^+$. We also show $m_F$ \textit{vs} $\tilde{U}$ for a set of $\alpha$ values in the upper panel of figure~\ref{fig3:staggered magnetism}. For finite $\alpha$, $m_F$ increases initially with increasing $\tilde{U}$, peaks at a given $\tilde{U_p}(\alpha)$ and then decreases with increasing  $\tilde{U}$. The $\tilde{U_p}(\alpha)$ shifts to lower $\tilde{U}$ with increasing $\alpha$, and $\tilde{U}_p(\alpha) \rightarrow  0$ for $\alpha \rightarrow 1$. In the strong coupling regime, the $m_F(\tilde{U})$ \textit{vs} $\tilde{U}$ curves merge together for all values of the inhomogeneity and approach zero asymptotically.
Further, we show the uniform magnetization $m_F$ for varying inhomogeneity at different Hubbard interactions in the lower panel of figure~\ref{fig3:staggered magnetism}. Below a given interaction strength, $m_F$ increases with increasing $\alpha$, but the uniform magnetization curve goes to an inflection point. The inflection point shifts to higher $\alpha$ with decreasing $U$. The inflection in the curve appears at $\alpha\rightarrow 1$ in the limit $U\rightarrow 0$ indicating a sharp crossover to ferromagnetic state in the Lieb lattice limit. Such magnetic behavior can be assigned to the flat band ferromagnetism. For the Lieb lattice limit ($\alpha\rightarrow 1$),  $B$ and $C$ sub-lattices (sites with flat bands) are polarized, with vanishing magnetization at $A$ sub-lattice, for infinitesimal strength of the interaction. Above the crossover interaction strength the curvature of staggered magnetization is positive and the magnetic behavior is determined by local interactions mainly. Emergence of such uniform magnetization is detrimental to the singlet $d_{x^2-y^2}$ pairing superconductivity defined in equation~\ref{eq:dwave}. We will discuss the influence of the inhomogeneity on the superconducting order in section~\ref{sec:dop}.

\subsection{Double occupancy}
\label{sec:double}
\begin{figure}
\includegraphics[scale=0.6]{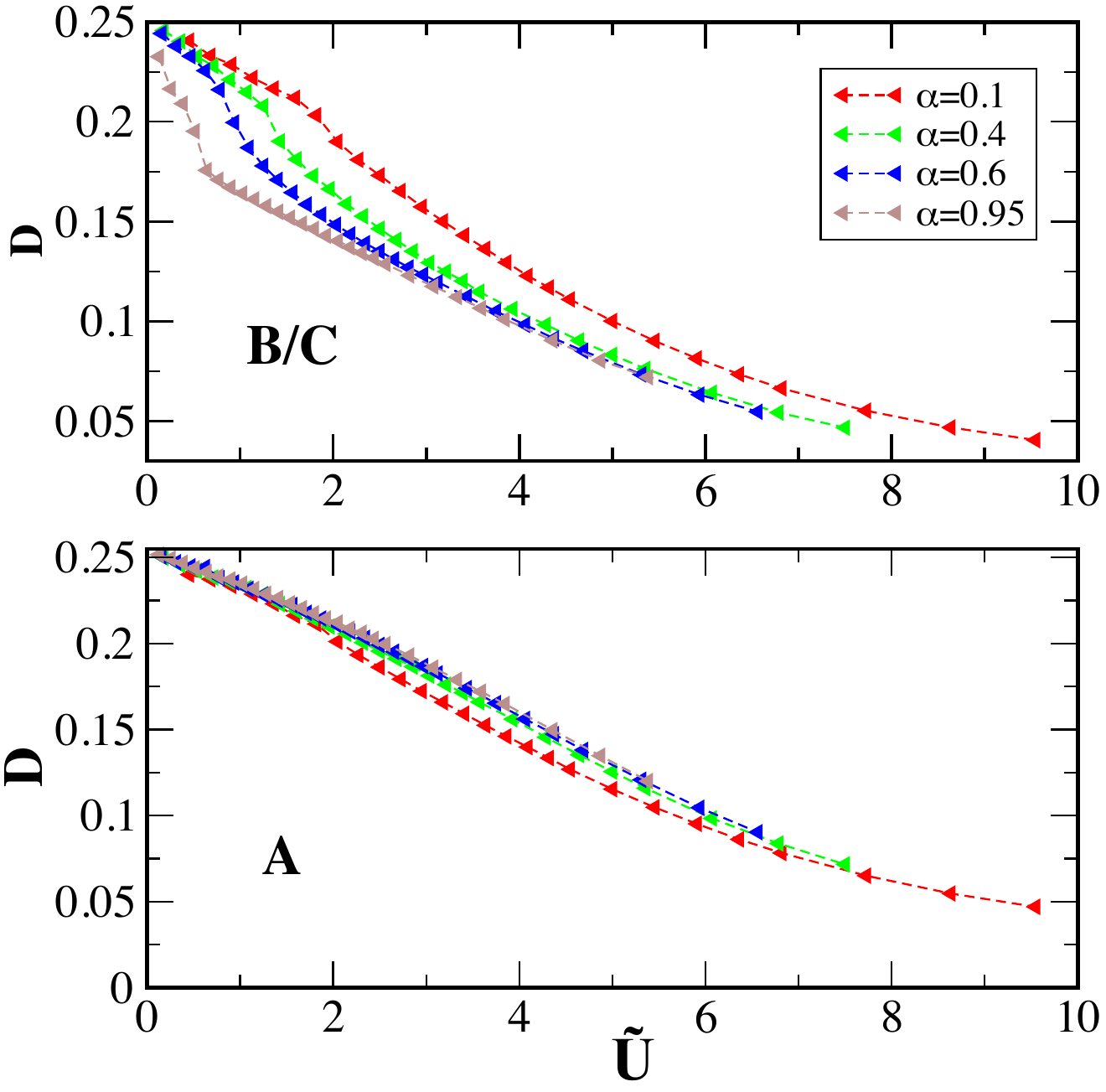}

\caption{Upper panel: Double occupancy of $B/C$ sites for varying interaction $\tilde{U}$ and different $\alpha$ at $\beta=1/T=20$. Lower Pane: Double occupancy of $A$ sites.}
\label{fig4:double occupancy}
\end{figure}

In this section, we study the interplay of the inhomogeneity and the Hubbard interaction in double occupancy at a given site, i.e. $\langle \hat{n}_{\uparrow}\hat{n}_{\downarrow} \rangle$ using RDMFT+CT-INT. Double occupancy can be a direct measure of local moment formation, $\langle m_z^2 \rangle=\langle (\hat{n}_\uparrow-\hat{n}_\downarrow)^2 \rangle=\langle (\hat{n}_\uparrow +\hat{n}_\downarrow-2 \ \hat{n}_\uparrow \hat{n}_\downarrow)\rangle= n -2\ \langle\hat{n}_{\uparrow}\hat{n}_{\downarrow}\rangle$. In the non-interacting limit, the up and down electrons are decoupled, $\langle \hat{n}_{\uparrow}\hat{n}_{\downarrow} \rangle=\langle \hat{n}_{\uparrow} \rangle\langle \hat{n}_\downarrow\rangle$. In figure~\ref{fig4:double occupancy}, we compare the spatially resolved double occupancy for different inhomogeneities $\alpha$ at small finite temperature $\beta=1/T=20$. The double occupancy of the $B/C$ sites is shown in the upper panel and that of the site $A$ is shown in the lower panel. At $B/C$ sites the double occupancy sharply decreases with increasing $\alpha$ for moderate values of $\tilde{U}$. The presence of a flat band favors single occupancy even for infinitesimal interactions as indicated by the sharp decrease of the double
occupancy. Kink in the double occupancy variation corresponds to the critical interaction, $U_c$, for the magnetic transition at given temperature. $U_c \rightarrow 0^+$ for Lieb lattice limit at zero temperature~\cite{PhysRevA.80.063622,PhysRevB.96.245127}. The double occupancy $D_{B/C}\rightarrow 0.1875$ for $\alpha \rightarrow 1$, $U \rightarrow 0^+$ and $T\rightarrow 0$. This limiting case can be explained as follows. The local magnetization at the $B/C$ sites at $T=0$ and $U\rightarrow 0^+$ is $0.5$ for the Lieb lattice at half-filling, i.e. the average number of particles per site is one~\cite{PhysRevA.80.063622}, and thus we can write
\begin{eqnarray}
n_{B/C\uparrow}+n_{B/C\downarrow}=1.0; \
n_{B/C\uparrow}-n_{B/C\downarrow}=0.5,
\end{eqnarray}
giving $\langle \hat{n}_{B/C\uparrow} \rangle = n_{B/C\uparrow}=0.75$ and $ \langle \hat{n}_{B/C\downarrow} \rangle = n_{B/C\downarrow}=0.25$, and thus $D_{B/C}=\langle \hat{n}_{B/C\uparrow}\rangle\langle \hat{n}_{B/C\downarrow} \rangle=0.1875$ in the $U\rightarrow 0^+$ and $T\rightarrow 0$ limit. In the strong coupling limit, for large $\tilde{U}$, double occupancy for different inhomogeneities coalesces and goes to zero asymptotically. At site $A$ the double occupancy coalesces to a single decreasing curve with varying $\tilde{U}$ for moderate to large inhomogeneity. 
\subsection{Non-Fermi liquid behavior}
\label{sec:nfl}
\begin{figure}
\includegraphics[scale=0.5]{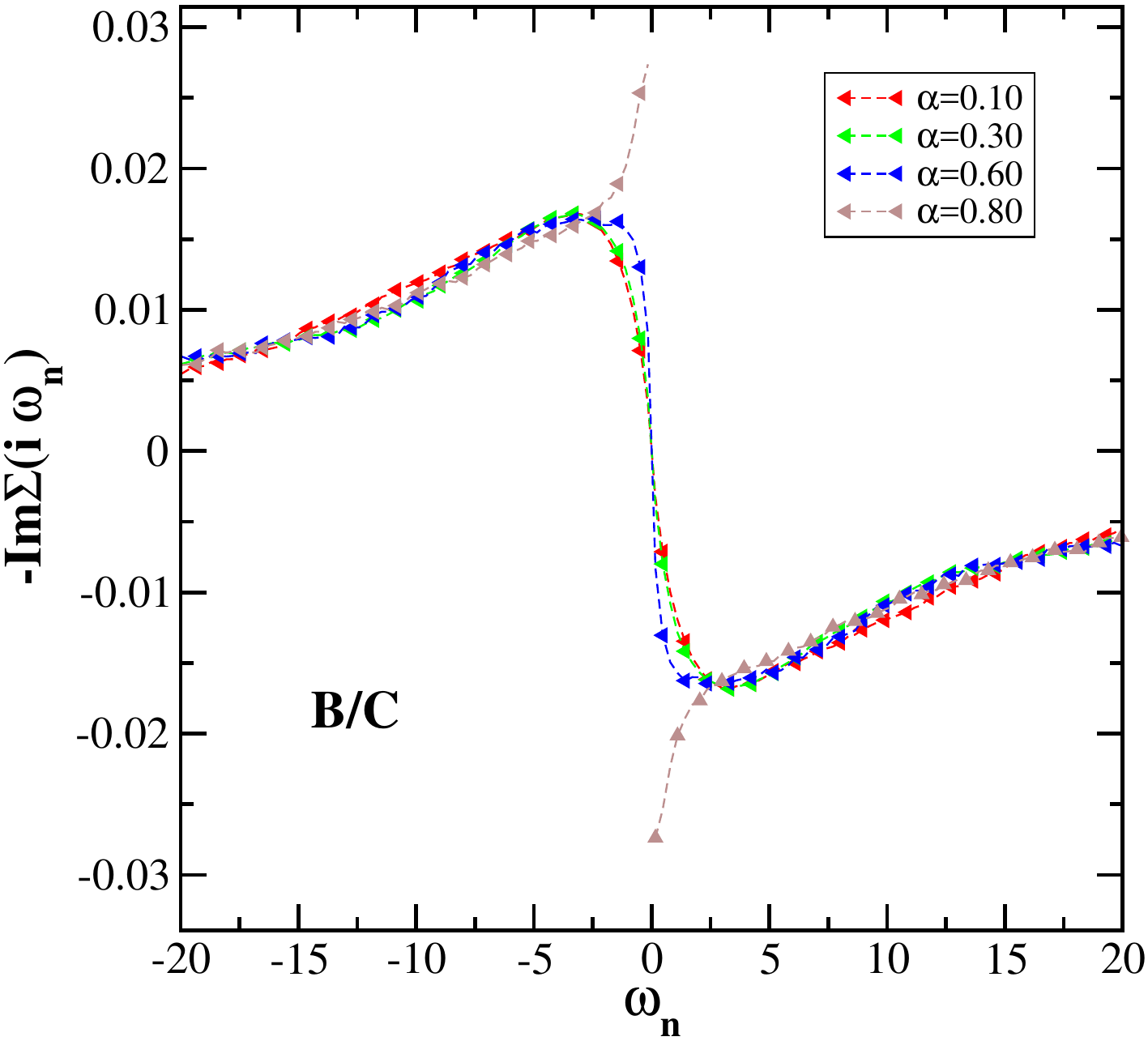}
\includegraphics[scale=0.5]{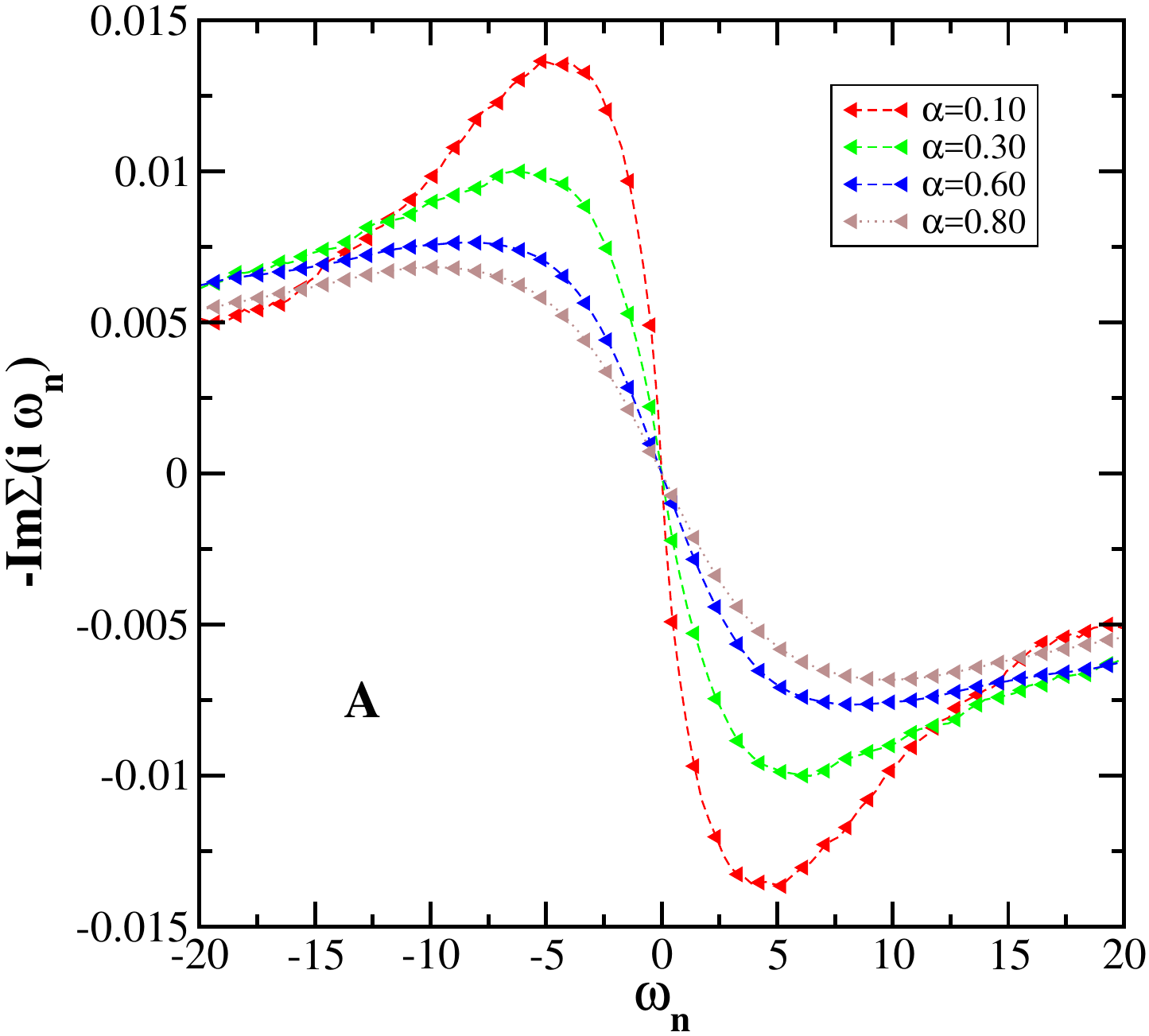}
\caption{Upper panel : Imaginary part of the local self-energy, i.e. $-\text{Im}\Sigma(i\omega_n)$ \textit{vs} Matsubara frequency $\omega_n$, for  site $B$ and $C$, for different $\alpha$, $T=0.05$ and $U=0.75$. For these parameters the system is in the non-magnetic metallic regime~\cite{0034-4885-68-10-R02}. Lower panel:  $-\text{Im}\Sigma(i\omega_n)$ vs Matsubara frequency $\omega_n$, for  site $A$ for the same parameter as upper panel.}
\label{fig5:non-Fermi liquid}
\end{figure}

We explore the effect of inhomogeneity on quasi-particle behavior in the weak coupling regime in the non-magnetic region at small finite temperatures using RDMFT+CT-INT. We find breakdown of the usual Fermi-liquid behavior occurs beyond a critical strength of the inhomogeneity, which is evident from the scattering rate, i.e. the imaginary part of the local self-energy, for different sites within the unit cell. There have been a few theoretical proposals for the origin of non-Fermi liquid behavior linked to the presence of singularities in the dispersion of the noninteracting part of the Hamiltonian~\cite{PhysRevLett.63.1996,PhysRevB.46.11798,PhysRevB.51.9253,refId0,PhysRevB.68.195101}. Non-Fermi liquids have also been observed within theories which include non-local correlations~\cite{PhysRevB.79.045133,PhysRevB.80.165126}. For a well defined Fermi liquid, the self-energy for low Matsubara frequencies $\omega_n$ can be written as
\begin{equation}
\Sigma(i\omega_n)\approx i\omega_n  a  +b,\label{eq9}
\end{equation}
where $a$ and $b$ are real constants. The quasi-particle weight $Z=m/m*$, where $m$ is the bare mass and $m*$ is the mass in the presence of many-body effects, can be defined in terms of the self-energy as
\begin{equation}
Z=\big(1-\frac{\partial \text{Im}\Sigma(i\omega_n)}{\partial \omega_n}|_{n = 0; \ T\rightarrow 0}\big)^{-1}\label{eq10}
\end{equation} 
and $0 < Z < 1 $ for the Fermi-liquid.
We observe the imaginary part of the self-energy at the lowest numerically calculated Matsubara frequency $\omega_0$ and at the next consecutive frequency $\omega_1$ and define ($a=|\text{Im}\Sigma(i\omega_0)| - |\text{Im}\Sigma(i\omega_1)| )$ such that $a < 0 $ signifies a Fermi-liquid while $a > 0$ is characteristic of a non-Fermi-liquid. In the upper panel of figure~\ref{fig5:non-Fermi liquid}, we show the imaginary part of the self-energy at $B/C$ for different inhomogeneities. For small to moderate values of the inhomogeneity, the system is a Fermi-liquid with $a < 0$ and well defined quasi-particle weight $Z$. For large inhomogeneity, say $\alpha=0.80$, the self-energy for the $B(C)$ sites, which carry the flat band, diverges for small frequencies $|\omega_n|$ and we observe non-Fermi-liquid behavior with $a > 0$ where quasi-particle weight cannot be well defined. In the lower panel of figure~\ref{fig5:non-Fermi liquid}, we show the self-energy for the $A$ site. The quasi-particle weight can be defined for all inhomogeneities since $a < 0$ although it increases with increasing $\alpha$.\\

\begin{figure}
\includegraphics[scale=0.6]{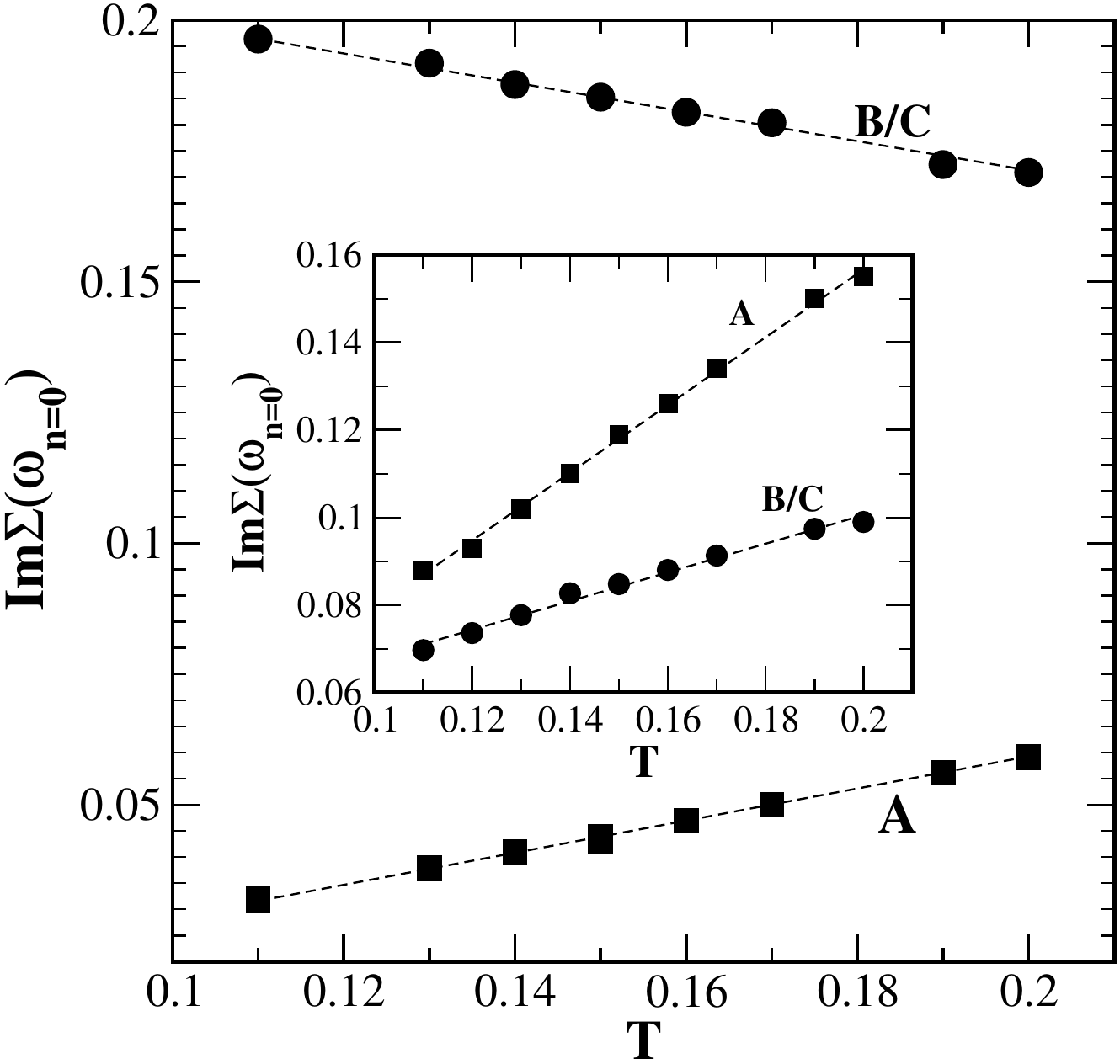}
\caption{In the main panel: $\text{Im}\Sigma(i\omega_{n=0})$ \textit{vs} $T$ at different sites for $\alpha=0.95$ and $U=2$. A similar plot is shown for $\alpha=0.4$ in the inset. For these parameters the system is in the non-magnetic metallic regime~\cite{0034-4885-68-10-R02}.}
\label{fig6:non-Fermi liquid_int}
\end{figure}

 \indent In the figure~\ref{fig6:non-Fermi liquid_int}, we present $\text{Im}\Sigma(i\omega_{n=0})$ which is an estimate of the inverse of the scattering time $\tau^{-1}\approx -\text{\Im}\Sigma(i\omega_{n=0})$. For Fermi-liquid behavior (conventional metallic behavior) the inverse of the scattering time, which is proportional to the resistivity, decreases with decreasing temperature. As shown in the main panel of figure~\ref{fig6:non-Fermi liquid_int}, we find breakdown of Fermi-liquid behavior as $\text{Im}\Sigma_{B/C}(i\omega_{n=0})$ increases with decreasing temperature for $\alpha \rightarrow 1$ and finite interaction $U=2.0$ while $\text{Im}\Sigma_{A}(i\omega_{n=0})$ decreases with decreasing temperature displaying Fermi-liquid behavior. In the inset of figure~\ref{fig6:non-Fermi liquid_int}, we show $\text{Im}\Sigma(i\omega_{n=0})$ \textit{vs} $T$ for moderate strength of the inhomogeneity, say $\alpha=0.4$. $\text{Im}\Sigma(i\omega_{n=0})$ decreases with decreasing temperature for both $B/C$ and $A$ sites and the system displays Fermi-liquid behavior.
  Non-Fermi liquid behavior in the presence of a flat band has been discussed previously for the multiband Hubbard model with repulsive interaction~\cite{PhysRevLett.115.156401, Hausoel2017}. Doping driven FL to NFL change has been found using DMFT calculations combined with first principles density functional theory~\cite{PhysRevLett.115.156401, Hausoel2017}. The origin of such NFL behavior was the nearly flat dispersion present in the given material. Also a multiorbital Hubbard model with orbital dependent hoppings has been studied in the context of orbital-selective~\cite{PhysRevB.72.205126} Mott transition, where the origin of NFL behavior is due to the lattice structure. In our study, we have systematically tuned the lattice model from dispersive to flat bands to show how the non-Fermi liquid behavior emerges. 
\subsection{Doped Hubbard model}
\label{sec:dop}
\begin{figure}
\includegraphics[scale=0.7]{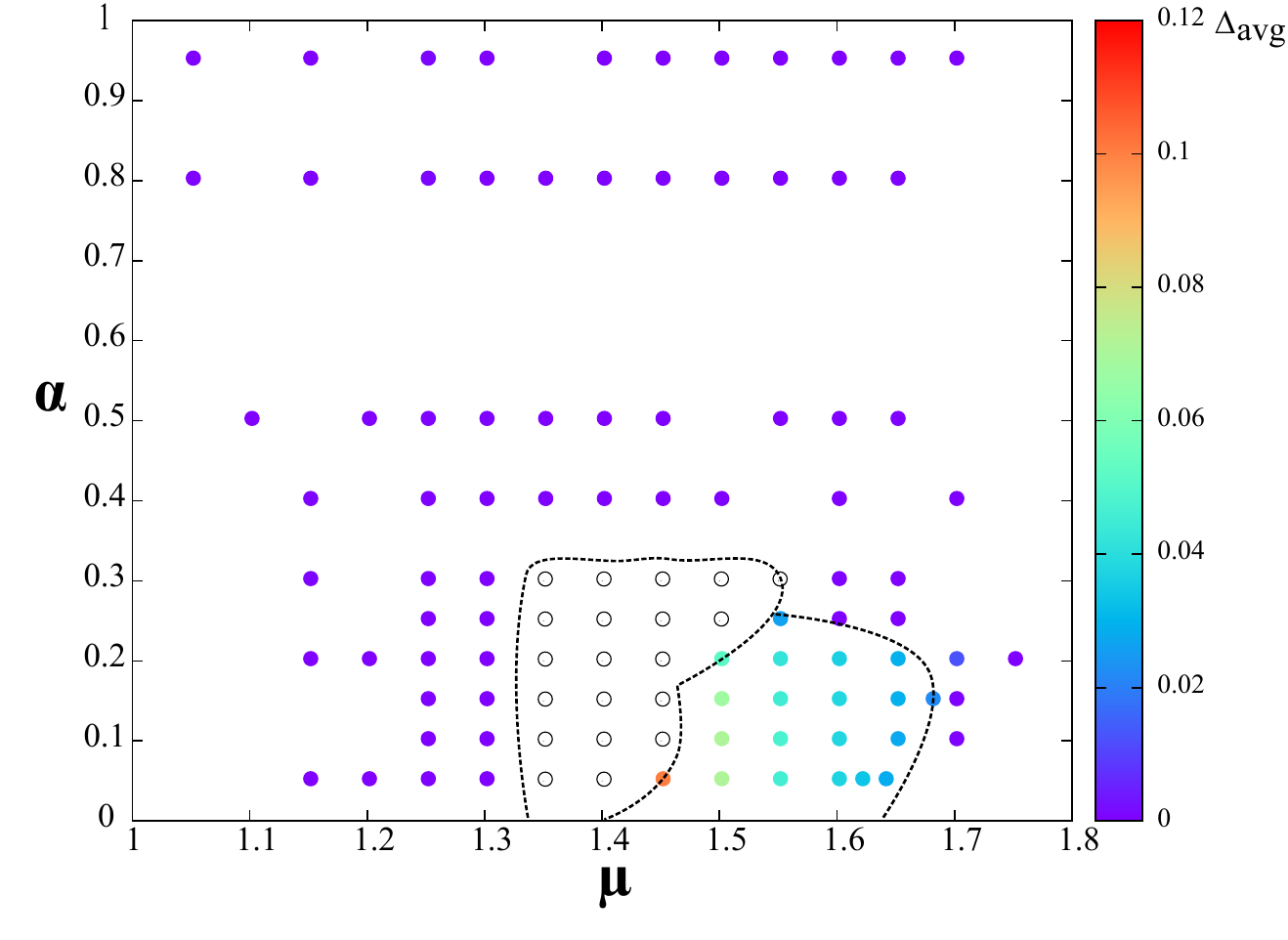}
\includegraphics[scale=0.7]{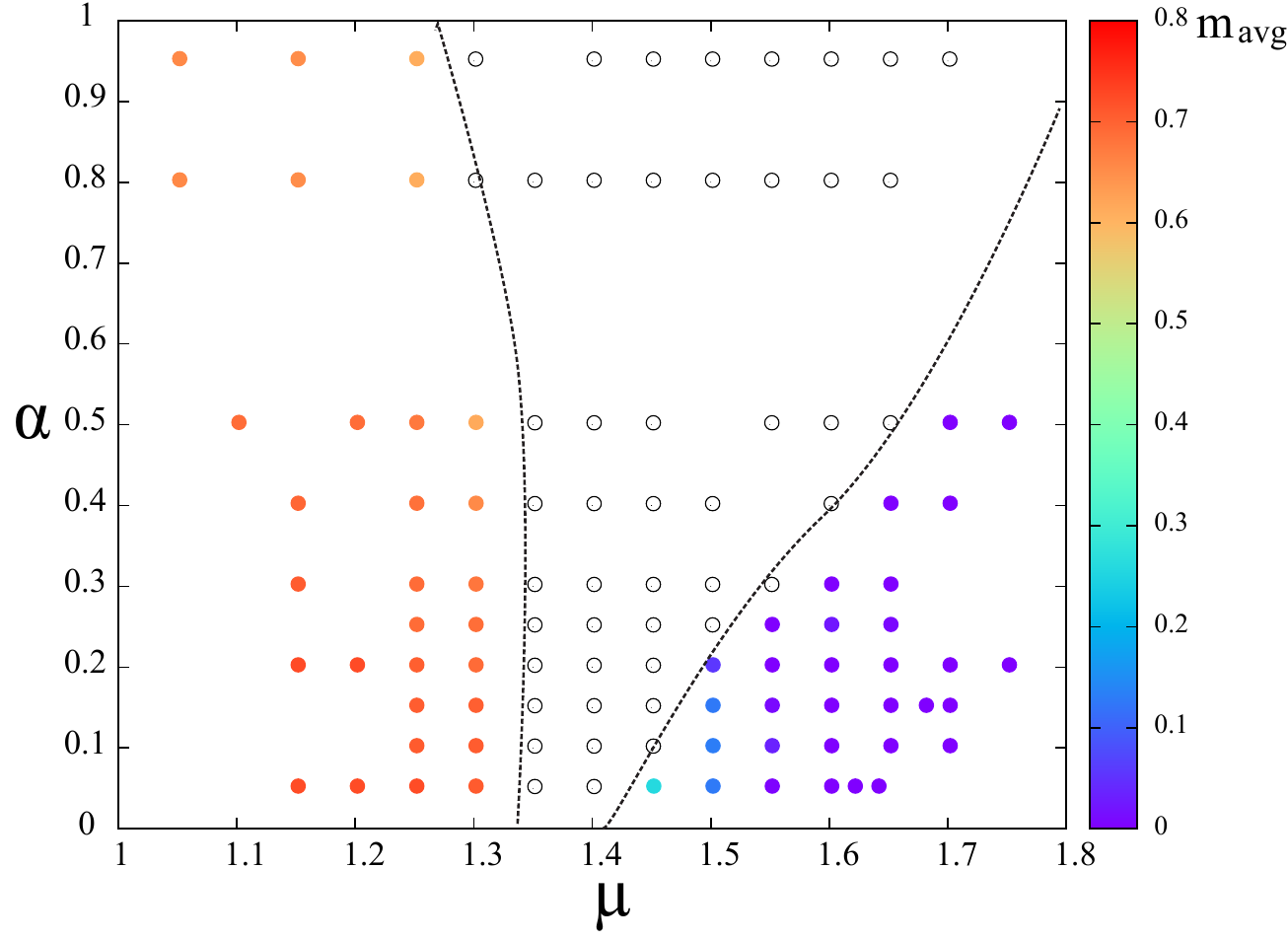}
\caption{Upper panel: Average dSC order parameter for set of inhomogeneity $\alpha$ and chemical potential $\mu$ for the two dimensional Hubbard model on the inhomogeneous square lattice with modulated hopping at $U=6.0$. The circles are the data points where we have carried out the plaquette DMFT+ED calculations. The dashed lines are guides to the eye separating different regions. The lines are only qualitative and do not actually correspond to a phase boundary. For solid circles, we obtain a converged DMFT solution while open white circles represent the data set for which DMFT solutions are finite and oscillatory. The color code assigned to the solid circles represent the magnitude of the dSC order parameter. Lower panel: Avergaed magnetic order as a function of inhomogeneity $\alpha$ and chemical potential $\mu$.}
\label{fig7:superconductivity phase}
\end{figure}

To explore the possible dSC in the presence of finite inhomogeneity away from half-filling, we have carried out cellular DMFT+ED calculations using a $2\times2$ cluster. Since the present choice of the inhomogeneity expands the unit cell by a factor of $2$ in each direction, the four-site plaquette actually  comprises a single unit cell of the model. This plaquette DMFT approximation is equivalent to the single site DMFT for a four band model in the sense that we get one impurity problem with four spin-degenerate orbitals. We uniformly dope the system by choosing a finite chemical potential $\mu$ independent of the lattice site in the unit cell. We allow breaking of the $SU(2)$ spin symmetry and thus long-range anti-ferromagnetic order.
We show the dSC order for different values of chemical potential $\mu$ and inhomogeneity $\alpha$ in the upper panel of figure~\ref{fig7:superconductivity phase}. We observe a region with finite dSC order parameter for moderate inhomogeneity, while dSC is not present for inhomogeneity $\alpha \geq 0.4$ for any finite $\mu$. We also observe a region where the dSC oder parameter is finite but non-convergent and oscillates with the DMFT iteration with a period longer than two iteration steps as shown by the circles. We also present the behavior of local magnetization averaged over the unit cell for different $\mu$ and $\alpha$. We obtain a magnetic to non-magnetic crossover going through a region with magnetic order oscillating with the DMFT iteration in the lower panel of figure~\ref{fig7:superconductivity phase}. In this case such oscillatory solution is observed for $0<\alpha< 1$ with varying $\mu$.  

\begin{figure}
\includegraphics[scale=0.55]{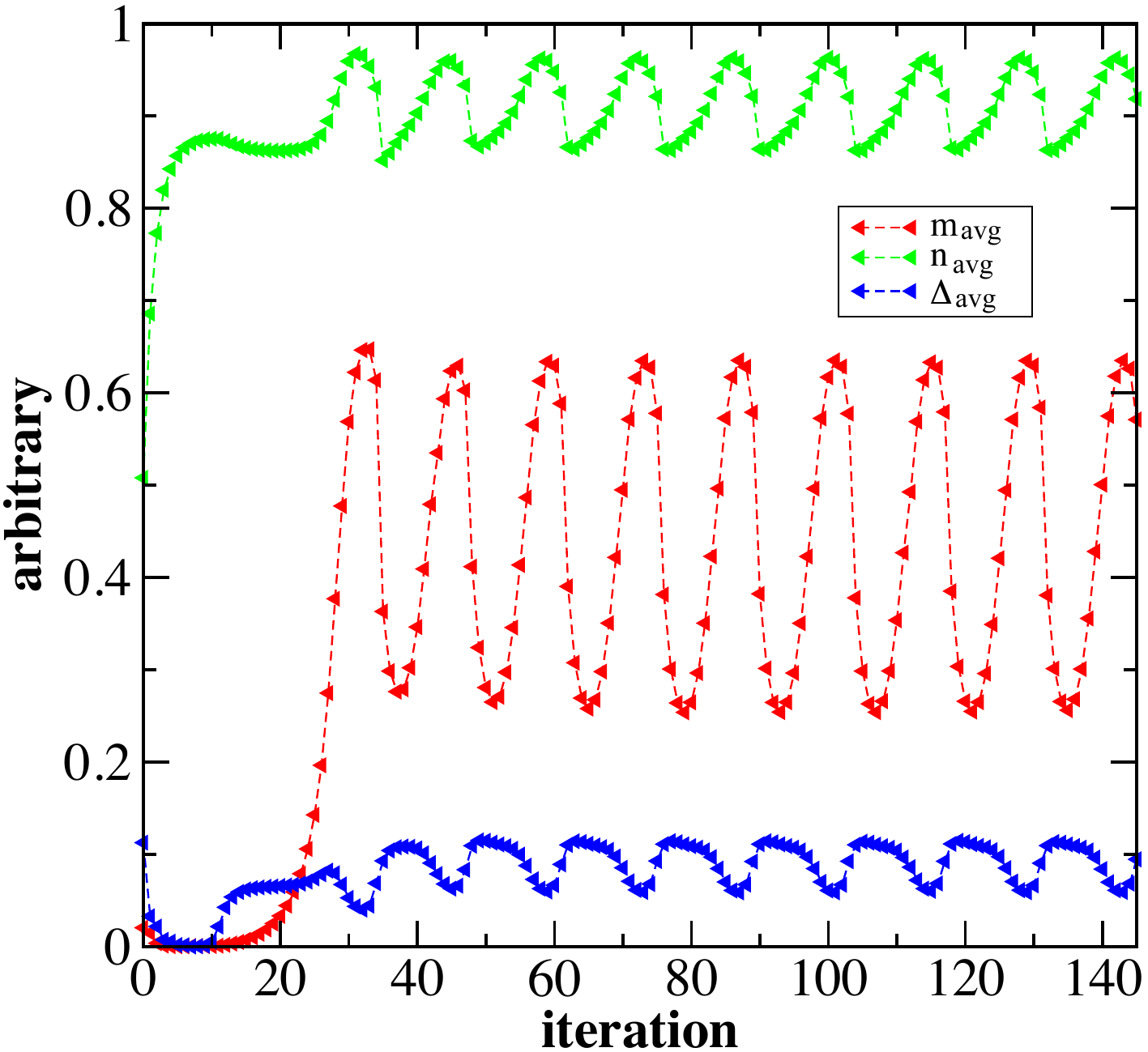}
\includegraphics[scale=0.55]{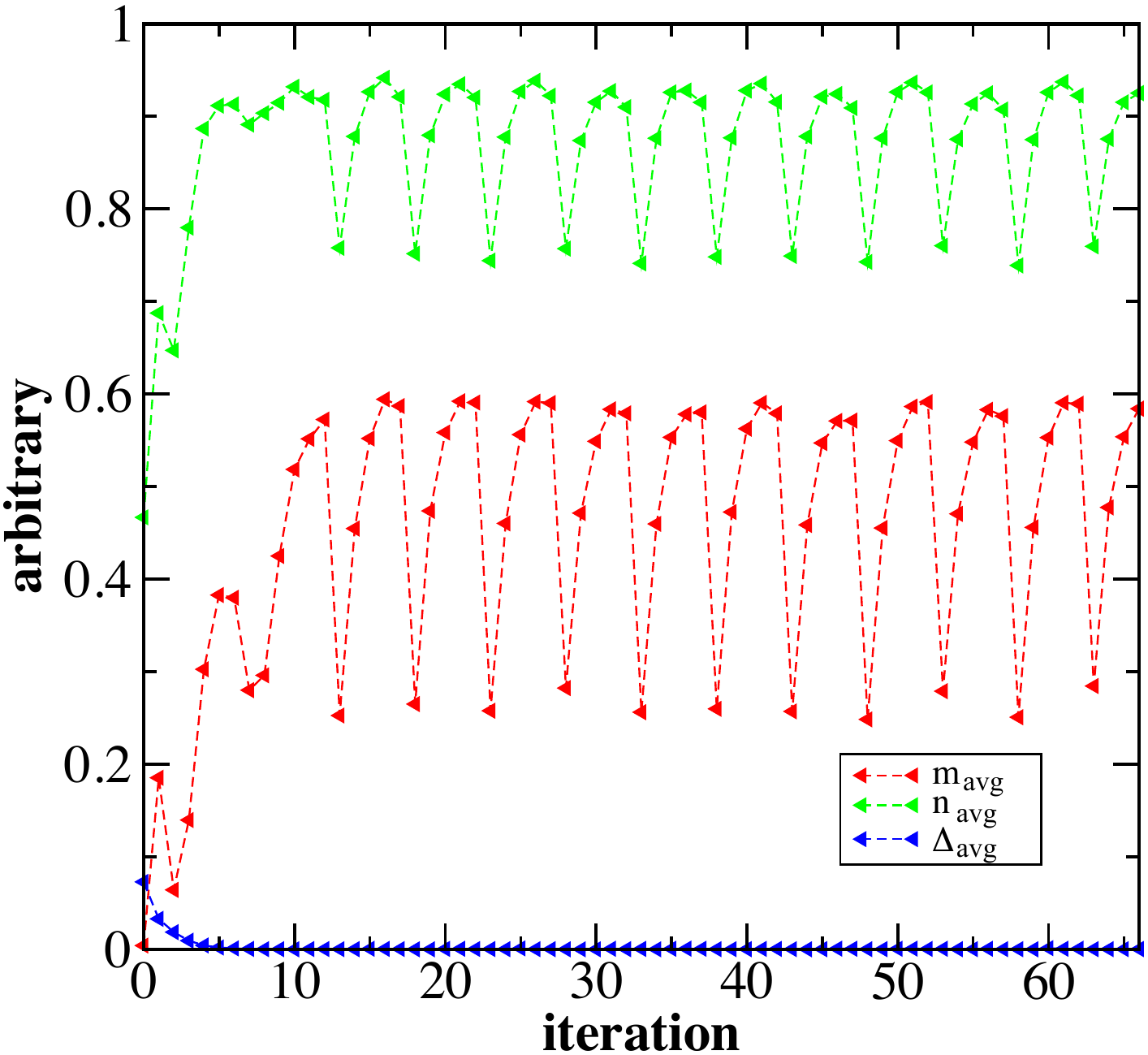}
\caption{Upper panel: The order parameters $m_{\text{avg}}$, $n_{\text{avg}}$ and $\Delta_{\text{avg}}$ for varying DMFT iterations for $\alpha=0.05$. All quantities show oscillatory behavior. Lower panel: Same order parameters for varying DMFT iterations for $\alpha=0.5$. Here $m_{\text{avg}}$ and $n_{\text{avg}}$ oscillate with the DMFT iteration, while $\Delta_{\text{avg}}$ converges to zero.}
\label{fig9:iteration}
\end{figure}

An example of the DMFT calculations in the region with oscillatory solutions can be seen in figure~\ref{fig9:iteration}. We show the results for two values of the inhomogeneity, i.e. $\alpha=0.05$ in the upper panel and $\alpha=0.5$ in the lower panel. For $\alpha=0.05$, the different order parameters such as dSC, magnetic and density order oscillate with the DMFT iteration with a period longer than two and a convergent solution cannot be achieved. Motivated by the observations in doped 2D homogeneous Hubbard model~\cite{PhysRevB.44.7455,PhysRevLett.74.186,PhysRevLett.80.2393,PhysRevB.60.5224}, such a behaviour has been interpreted as indication that an incommensurate spin density wave is the proper state~\cite{1367-2630-11-8-083022,PhysRevB.89.155134}
 and consequently calculations do not converge in this parameter region. Although there is no direct mathematical foundation for such an interpretation, we have previously reported presence of spatially non-uniform magnetic and charge order coexisting with dSC using an extended plaquette DMFT approximation for the canonical 2D Hubbard model. In that case calculations were carried out for unit cells with a large number of sites by taking one-dimensional slices of the lattice~\cite{vanhala2017dynamical}. There, incommensurate orders coexisting with dSC were reported, such as the spin density wave coexisting with inhomogeneous dSC of wavelength $12$ plaquettes which was found to have the lowest energy for $\mu=1.40$ and $U=6.0$. Such spatially non-uniform SDW orders reported in several recent works~\cite{,PhysRevB.89.155134,Zheng1155,salomon2018} brace the interpretation. The oscillatory solutions obtained using DMFT can be made to converge using different mixing techniques, but this is likely to lead to a metastable solution given that a long wavelength SDW is not allowed for the simple plaquette DMFT approximation. For moderate inhomogeneity $\alpha=0.5$ shown in the lower panel of figure~\ref{fig9:iteration}, we observe the oscillations only for the magnetic and density orders while the superconducting order converges to $\Delta_{\text{avg}}=0$. This behavior prevails for moderate to large inhomogeneity. 
 It is also possible that other types of orders such as phase separation could exist in the region where non-convergent solutions are found~\cite{PhysRevB.74.085104}. A typical sign for phase separation is a first order jump in the density with tuning $\mu$ \cite{1367-2630-11-8-083022}, and such sensitivity to $\mu$ is also associated with the region of oscillatory solutions.

\begin{figure}
\includegraphics[scale=0.55]{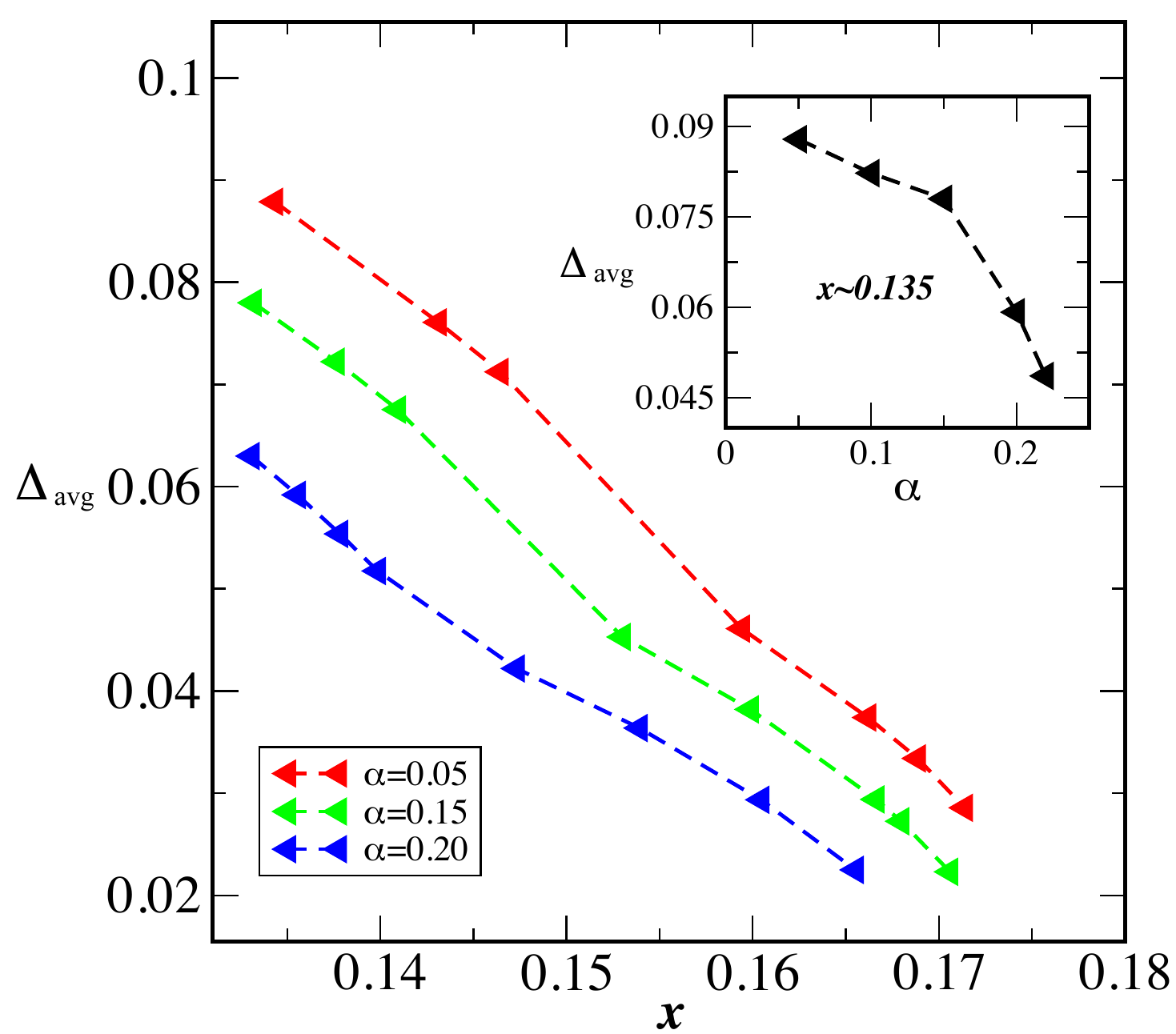}

\caption{Main panel: Uniform dSC order parameter for converged DMFT solutions, $\Delta_{\text{avg}}$, vs doping $x$ for different values of $\alpha$ and $U=6.0$ at $T=0$. Uniform dSC order monotonically decreases with increasing $x$. Magnitude of dSC is smaller for larger inhomogeneity at given $x$ and is zero for $\alpha \ge 0.25$. In the inset: Uniform dSC order parameter for varying $\alpha$ for given $x=0.135$. }
\label{fig9:superconductivity}
\end{figure}

 The results for the uniform dSC order parameter corresponding to converged DMFT solutions, for several values
of inhomogeneity $\alpha$ displayed in the figure~\ref{fig9:superconductivity},
exhibit interesting features. We find that the strength of dSC decreases monotonically as a function of $\alpha$ over the entire doping range. Our findings complement the results of previous studies of interplay of lattice inhomogeneity and interactions in the context of dSC on the checkerboard lattice using CDMFT~\cite{PhysRevB.84.054545}. CDMFT calculations show a monotonic decrease in the dSC order with inhomogeneity i.e. the ratio of the inter-plaquette to intra-plaquette hopping. Dynamical cluster approximation (DCA) finds monotonic decrease of the critical temperature with strength of the inhomogeneity~\cite{PhysRevB.78.020504}. In contrast, DQMC calculations for similar inhomogeneity pattern find an optimal value for which the pair vertex is most attractive~\cite{PhysRevB.90.075121}. In both approaches the dSC order eventually vanishes for large inhomogeneity. In the present study, dSC is completely destroyed for $\alpha \ge 0.25$.  A few other patterns of inhomogeneity where an onsite potential of one fourth of the lattice sites of the square lattice is raised by an amount $V_0$ such that in the limit $V_0 \rightarrow \infty$, the lattice maps onto the ``Lieb lattice'' Hamiltonian have been studied~\cite{PhysRevB.90.075121}. It has been found that this kind of inhomogeneity rapidly, and monotonically, suppresses the dSC pairing. 
\section{Conclusions}
To understand the spatial non-uniformity of the
various order parameters in systems ranging from real materials to cold atom systems, Hubbard Hamiltonians with different inhomogeneity patterns have been proposed. The pattern of inhomogeneity explored in the present work leads to the Lieb lattice geometry as a limiting case. Importantly, this allows the study of the effect of an emerging flat band singularity. We have applied RDMFT to explore the influence of inhomogeneity on different physical properties at half-filling and finite Hubbard interactions. The inhomogeneity changes the magnetic behavior of the system, interpolating between the square lattice and Lieb lattice cases. Below a given interaction strength, the uniform magnetization displays a sharp crossover to a ferromagnetic state with increasing the inhomogeneity. There is an associated inflection point in the uniform magnetization vs the inhomogeneity parameter, with the sharp crossover. Such a behavior is due to a flat-band dispersion appearing when tuning of the inhomogeneity. We also observe a breakdown of Fermi-liquid behavior when the inhomogeneity is increased, signalled by the inverse scattering time defined by the local self-energy.
  
 To capture the non-local d-wave superconductor (dSC) order parameter away from half-filling, we employ cellular dynamical mean field theory (CDMFT) combined with an ED impurity solver for a cluster of four sites $(2\times2)$. For a range of doping values we observe oscillatory behaviour in the DMFT iteration, which we tentatively associate with incommensurate spin-density-wave order. For small inhomogeneity the system displays uniform dSC and also dSC coexisting with the incommensurate order depending on the chemical potential. We find suppression of the dSC order parameter for moderate to large inhomogeneity, while the oscillatory solutions associated with incommensurate order persist for all finite values of the inhomogeneity. The presence of incommensurate order coexisting with dSC in the homogeneous case is in accordance with recent findings~\cite{vanhala2017dynamical,PhysRevB.98.205132}, although further work would be needed to determine the actual wavelength and other properties of the spin-density-wave.

Our findings can be relevant to ultracold gas experiments, where the simple two-dimensional Hubbard model \cite{Sherson2010,Tarruell2012,Greif953,Cheuk1260,PhysRevLett.116.235301,PhysRevLett.116.175301,Brown1385} as well as different inhomogeneity
 patterns and lattice geometries \cite{Tarruell2012,Struck996,PhysRevLett.115.115303,PhysRevLett.119.230403} have been realized. Experimentally, the geometry of an optical
lattice can be determined by the spatial arrangement of the laser beams, and the tunnelling of the trapped atoms within the lattice is
then tuned via the laser amplitudes~\cite{doi:10.1146/annurev-conmatphys-070909-104059}. Spin correlations displaying antiferromagnetic behavior have been observed using Bragg scattering~\cite{Hart2015} and fermionic microscopes~\cite{Parsons1253,mazurenko2017cold}. Using these techniques, it could be possible to also study magnetism in optical Lieb lattices populated with fermionic atoms \cite{taie2017spatial,Leykam2018}. Our results show how an imperfect, quasi-flat band affects the double occupancy and magnetization, and could thus aid interpretation of such experimental results. It could also be possible to experimentally engineer the exact model that we have proposed here. A square-to-Lieb-lattice crossover could be studied by tuning the laser amplitudes in the configuration used in previous experiments \cite{Taiee1500854}, although the corresponding tight-binding lattice will also include on-site potential contributions on the D-sites. Nevertheless, this is perhaps the easiest way to study a tunable flat-band within ultracold gas systems.

\section*{Acknowledgments} 

This work was supported by the Academy of Finland through its Centers of Excellence Programme (2012-2017) and under Project Nos.284621, 303351 and 307419, by the European Research Council under ERC-2013-AdG-340748-CODE and ERC-2017-COG-771891-QSIMCORR and by the Deutsche Forschungsgemeinschaft (DFG, German Research Foundation) under Germanys Excellence Strategy EXC-2111 390814868. Computing resources were provided by CSC -- the Finnish IT Centre for Science.

\bibliographystyle{apsrev4-1} 
\bibliography{apssamp}
\end{document}